\documentclass{article}
\usepackage{iclr2027_conference,times}
\usepackage[T1]{fontenc}
\usepackage[utf8]{inputenc}
\usepackage{textcomp}
\usepackage{amsmath,amssymb}
\usepackage{graphicx}
\usepackage{booktabs,array,calc}
\usepackage{placeins}

\usepackage{microtype}
\usepackage{xurl}
\usepackage{hyperref}
\hypersetup{hidelinks,pdfauthor={Felix Desyatirikov}}
\DeclareUnicodeCharacter{2423}{\textvisiblespace}
\DeclareUnicodeCharacter{2264}{\ensuremath{\leq}}
\providecommand{\tightlist}{\setlength{\itemsep}{0pt}\setlength{\parskip}{0pt}}
\makeatletter
\newsavebox\pandoc@box
\newcommand*\pandocbounded[1]{%
  \sbox\pandoc@box{#1}%
  \Gscale@div\@tempa{\textheight}{\dimexpr\ht\pandoc@box+\dp\pandoc@box\relax}%
  \Gscale@div\@tempb{\linewidth}{\wd\pandoc@box}%
  \ifdim\@tempb\p@<\@tempa\p@\let\@tempa\@tempb\fi%
  \ifdim\@tempa\p@<\p@\scalebox{\@tempa}{\usebox\pandoc@box}%
  \else\usebox{\pandoc@box}\fi%
}
\makeatother
\iclrfinalcopy
\author{Felix Desyatirikov\\ITMO University, St. Petersburg, Russia}

\title{Watermark Claims Under Test: Lessons from Auditing a Production LLM}
\hypersetup{pdftitle={Watermark Claims Under Test: Lessons from Auditing a Production LLM}}
\begin{document}
\maketitle
\pagestyle{plain}
\begin{abstract}
Can we tell whether an LLM is watermarked without knowing its secret key? Published black-box tests suggest that carefully designed prompts can reveal the statistical footprint of a watermark. This paper examines what happens when those tests move from controlled reference implementations to a deployed model whose watermarking parameters are undisclosed. We present an audit of the Red-Green watermark presence test on Claude. Across ten initial configurations, the published test never rejects on a model the provider states is watermarked, despite detecting a local reference at comparable sample sizes. Controlled experiments expose a failure mode: including the prompt in the watermark’s repeated-context history suppresses marking exactly where the test measures it. The watermark remains active, but the probe becomes blind.

Attempts to repair the probe reveal further confounds. Constructed contexts can trigger native digit-to-word preferences, while comparisons within a single response can measure how the model plans its answer. Tokenization changes which contexts the probe actually controls. These findings show why validation on a surrogate model does not establish validity on the model under investigation.

A redesigned probe finds a substantial context-dependent shift on Claude Fable that is absent on two sibling controls. However, the pre-registered decision remains inconclusive because selective refusals violate the protocol, and the effect’s watermark origin is unresolved.

The paper focuses on the methodological lessons: how to design informative controls, distinguish statistical rejection from evidence of watermarking and interpret a null without treating it as proof of absence. We provide researchers and practitioners working on LLM evaluation, watermarking, and provenance with an audit checklist grounded in concrete successes and failures. Code, prompts, and parsed responses are available for reproducing the analyses.
\end{abstract}
\section{Introduction}\label{sec:1}

Watermarking offers a machine-detectable signal of generated text,
addressing provenance concerns associated with foundation models
\citep{bommasani2021} and motivating regulatory scrutiny
\citep{eu2024,nemecek2026}. Keyed detection asks
whether a text carries that signal. A different question matters to an
external auditor or an adversary preparing a watermark-stealing attack
\citep{jovanovic2024}: can someone without the key determine whether a
model is watermarked at all?

\textbf{\boldmath Related work.} Sampling-time schemes alter
token selection; weight-level schemes can learn their signature through
distillation \citep{gu2024}; black-box detectability has been debated
for distribution-preserving schemes \citep{hu2024,christ2024,piet2023}.
\citet{liu2025} elicited watermark biases with crafted prompts.
\citet{gloaguen2025} developed presence tests for Red-Green
\citep{kirchenbauer2023,kirchenbauer2024}, Fixed-Sampling
\citep{kuditipudi2024} and Cache-Augmented schemes
\citep{hu2024,wu2024}; their Red-Green test detected
SynthID-Text \citep{dathathri2024} on models they watermarked
themselves, and vendor APIs, none stated to be watermarked, returned
nulls. The difference from this paper is the ground truth: they validate
the test on watermarks the tester controls, whereas we ask what it
establishes on a deployment whose watermark the tester does not control,
and find that the answer turns on assumptions their analysis leaves
implicit. A separate line measures what a watermark does to the text:
downstream utility drops \citep{ajith2024}, instruction following
degrades at matched strength \citep{tu2024}, and alignment shifts in
model-specific directions \citep{verma2025}. Such side effects are
candidate presence signals, but each needs the same model without its
watermark as a baseline, which an external auditor lacks.

Anthropic's 2026 announcement creates the test case:
models released after 2 August carry a version of SynthID-Text, its
parameters are undisclosed, and its detection API was unavailable to us.
Our audit uses only prompts and responses on the hosted models, with
controlled experiments on an open reference whose watermark status is
known. We do not propose a new general detector; the contribution is
what a published presence test can and cannot establish about a
deployment:

\begin{enumerate}
\def\labelenumi{\arabic{enumi}.}
\tightlist
\item
  \textbf{\boldmath A mechanism the published test cannot
  see.} Seeding the watermark's repeated-context history from the
  prompt, a one-line implementation choice absent from the reference
  code, leaves the watermark active while every probe that places the
  context in the prompt measures an unmarked continuation: a
  counterexample with known ground truth, not a claim about Claude's
  implementation.
\item
  \textbf{\boldmath Two confounds that a clean surrogate
  cannot reveal.} Naming a digit in the instruction creates native
  digit-to-word preferences, and a two-line request changes the model's
  first choice before any context can repeat. Both redesigns pass the
  local control and fail the deployed control, so same-family controls
  are necessary.
\item
  \textbf{\boldmath A pre-registered audit with explicit
  attribution limits.} A long model-assembled context, scored within
  each prefix, produces a context-dependent shift on Fable (continuous
  range 0.57 against 0.07 and 0.12 on two siblings) that survives
  estimator, subsampling and low-refusal checks. The primary statistic
  rejects, but the planned decision is inconclusive because refusals are
  concentrated by context. When the provider later marked the siblings,
  one of them acquired a shift of the same size with the same column
  pattern and no refusals, which favours a keyed watermark over the two
  alternatives we keep open.
\item
  \textbf{\boldmath The conditions under which the test
  works}: a power floor set by watermark strength rather than sample
  size, the entropy range in which detection is possible, how
  tokenization fixes the controlled context, and how much column
  structure an unwatermarked reference shows under a long assembled
  context, summarised as an audit checklist.
\end{enumerate}

Section~\ref{sec:2} defines the test and the
protocol, Section~\ref{sec:3} the controlled experiments on the
reference, Section~\ref{sec:4} the deployment audit and
Section~\ref{sec:5} the checklist and the limits; the appendices give
definitions, prompts, every grid and the deviations from the plans.
Code, prompts and stored responses are in the anonymous supplementary
archive; the public repository link is omitted for review.

\section{Test and audit protocol}\label{sec:2}

\subsection{What the presence test measures}\label{sec:2.1}

SynthID-Text biases sampling through a keyed tournament whose outcome
depends on the preceding \(h\) tokens. For fixed keys, a context
therefore gives a stable direction of preference among candidate words.
It also applies \emph{repeated-context masking}: after a context has
occurred, a subsequent occurrence is sampled without the watermark. The
reference \texttt{transformers} processor starts generation with an
empty context history; it does not scan the prompt. Our prompt-aware
variant seeds that history from the prompt and initializes the rolling
context from its suffix; the watermark keys and sampling rule are
unchanged (Appendix~\ref{app:A.6}).

The Red-Green probe asks the model to reproduce a sentence consisting of
a prefix \(t_1\) and a context \(t_2\), then choose one of four words.
Ten prefixes and nine contexts form ninety cells, sampled independently.
Contexts initially consist of one digit repeated \(H\) times. The
preferred word \(x^*\) maximises the summed logit over cells. Its
cellwise logits form a matrix \(L\), with prefixes as rows and contexts
as columns. After subtracting each row's median, cells below
\(-1.96\hat\sigma\) are flagged red and cells above \(+1.96\hat\sigma\)
green. Writing \(R_j\) and \(G_j\) for the two flag counts in column
\(j\), the released code computes

\[S(L)=\max(\max_j R_j,\max_j G_j)-\max(\min_j R_j,\min_j G_j).\]

An aligned context bias can make \(S\) large. The published null
permutes all cells, whereas our \emph{within-prefix null} permutes
contexts separately in each row, preserving native prefix differences.
Before the logit, the redesigned score clips cell rates to
\([1/(2n),1-1/(2n)]\), where \(n\) is the cell's parsed sample size; the
published score adds one pseudo-count per word. Empty cells receive
their row median in both observed and permuted grids. Full definitions
are in Appendix~\ref{app:A}. Within-prefix permutation assumes
exchangeability of context-labelled cell logits within each row. Unequal
sample sizes, native digit preferences and selective refusals can
violate this assumption; row centring alone does not establish it. A
rejection is therefore not a calibrated test of watermark absence in an
arbitrary deployment.

Two estimators matter. The paper's \(\hat\sigma\) is the square root of
the median row variance; the released code uses the median column
standard deviation of uncentred logits. Within-row permutation preserves
the former but can change the latter. We report both, and treat the
planned code estimator as primary rather than choosing retrospectively.
Throughout, \(p_{99}\) is the 99\% Clopper-Pearson upper bound for a
permutation \(p\)-value. We use 10 000 permutations unless stated
otherwise.

Interpretation requires an uncertain choice in each cell, a watermark
context contained in the controlled suffix, sufficiently stable keys, no
native context effect that mimics the watermark, and marking active at
the measured position. A small pool of keys can preserve an average bias
\citep[App. F.2]{gloaguen2025}. Native preferences and selective
missingness must therefore be tested, not absorbed into a watermark
interpretation.

\subsection{Four probes and the tokenizer}\label{sec:2.2}

The \emph{echo} probe spells out the context in both the instruction and
format example. The \emph{described} probe instead asks for a run such
as twelve copies of digit 3, keeping the compact run out of the prompt
but making the digit salient. The \emph{masking differential} requests
two lines and measures the second choice: type A repeats line one's
context and type B changes it. Describing both runs avoids prompt
masking, and sharing a request avoids cross-request key rotation. Its
statistic \(T\) sums type-by-word chi-square statistics over contexts;
the null permutes type labels within context. A \texttt{mixed} variant
repeats only the relevant trailing window, rather than the entire first
number. Full prompts and the differential statistic are in Appendices
\ref{app:F} and \ref{app:A}.

The final \emph{spaced} probe shows an arbitrary number with spaces
between digits and asks the model to write it compactly. Its nine
contexts differ only in their final digit. This is intended to create a
new output context without naming that digit in the instruction. Qwen
encodes digits individually; Claude's token-count endpoint supports
three-digit groups and separate spaces around the run. Under that
inferred segmentation, forty digits cover sixteen tokens before the
context reaches the prefix. A one-token context sees only the constant
trailing space and is invisible to the column statistic. The inference
and its limitations are detailed in Appendix~\ref{app:B}.

\subsection{Models, sampling and safeguards}\label{sec:2.3}

The local reference is Qwen2.5-3B-Instruct with stock SynthID-Text, no
watermark, or our prompt-aware variant. Defaults are nine tournament
keys, \(h=4\) and ten samples per cell. Locally, we generate one
response per prompt, locate the choice after its last digit-token run,
renormalise the processor's distribution over the four words' first
tokens, and sample it. This simulates sampling noise conditional on one
generated response; it does not verify the requested number, which
limits validation of the spaced probe.
Section~\ref{sec:3.3} repeats the design checks on
Qwen3-8B-Instruct with the number verified.

Hosted runs span 6 to 13 September 2026. Fable
(\texttt{claude-fable-5-1}) is stated to be watermarked. Haiku
(\texttt{claude-haiku-4-5}) and Sonnet (\texttt{claude-sonnet-5}),
released before the announced cutoff, serve as presumably unmarked
controls. These labels rest on the provider's statements, which as of 23
September still list both controls without a text watermark, not on
keyed verification (Appendix~\ref{app:A.3}). By 8
November 2026 the same table listed both as marked, and
Section~\ref{sec:4.4} re-queries them. Context length, key schedule and
masking rules are undisclosed (Appendix~\ref{app:A}). We sample ordinary
API completions using provider defaults, synchronously or in batches,
without requesting prompt caching. A request nonce precedes the measured
context. Refusals are dropped without a fallback model; response parsing
and the missing usage records are described in Appendices \ref{app:A},
\ref{app:G} and \ref{app:I}.

We select word lists by within-cell entropy because pooled balance can
hide deterministic choices. A run with mean per-cell top-word share
above 0.9 is uninformative. Initial hosted runs use five to thirteen
samples per cell, differential runs fifty-two, and the final audit up to
sixty-five. Only the statistic and decision rule fixed in the
corresponding plan are confirmatory; alternative scores and later
robustness checks cannot replace that decision.

\section{Controlled failures of the probe}\label{sec:3}

\subsection{A watermark that the test cannot see}\label{sec:3.1}

The published probe detects stock SynthID-Text and stays clean on the
local control. Table~\ref{tab:1} compares three probes on the same model
weights. The prompt-aware initialization blinds the echo probe: the
model copies a context that the processor now treats as already seen, so
the measured continuation is unmarked. Additional samples cannot recover
a watermark signal at that position. This establishes a failure
mechanism on the reference; it does not identify Claude's
implementation.

\begin{table}[tbp]
\centering
\caption{Local probe comparison on Qwen2.5-3B, ten samples per
cell, \(H=h=4\), collected separately from the baseline in
Appendix~\ref{app:C}. Entries use the code's estimator; \(p\) is the
99\% upper bound from 10 000 permutations for \(S\) and 500 for \(T\),
whose smallest attainable bound is 0.011. The described probe uses the
mangoes list, with eight or nine empty cells filled by row medians; the
other probes use the peaches list. Bold entries show echo blindness and
the described probe's false positive.}\label{tab:1}
\vspace{\baselineskip}
\begin{tabular}{@{}
  >{\raggedright\arraybackslash}p{(\linewidth - 6\tabcolsep) * \real{0.2255}}
  >{\raggedright\arraybackslash}p{(\linewidth - 6\tabcolsep) * \real{0.2549}}
  >{\raggedright\arraybackslash}p{(\linewidth - 6\tabcolsep) * \real{0.2647}}
  >{\raggedright\arraybackslash}p{(\linewidth - 6\tabcolsep) * \real{0.2549}}@{}}
\toprule
\begin{minipage}[b]{\linewidth}\raggedright
probe
\end{minipage} & \begin{minipage}[b]{\linewidth}\raggedright
SynthID
\end{minipage} & \begin{minipage}[b]{\linewidth}\raggedright
control
\end{minipage} & \begin{minipage}[b]{\linewidth}\raggedright
SynthID, prompt-aware
\end{minipage} \\
\midrule
echo & \(S = 9\), \(p = 0.0005\) & \(S = 1\), \(p = 1.0\) &
\textbf{\boldmath \(S = 1\), \(p = 1.0\)} \\
described & \(S = 8\), \(p = 0.0005\) & \textbf{\boldmath \(S = 6\),
\(p = 0.0007\)} & \(S = 8\), \(p = 0.0005\) \\
masking differential & \(T = 35.6\), \(p = 0.011\) & \(T = 8.8\),
\(p = 0.30\) & \(T = 28.5\), \(p = 0.011\) \\
\bottomrule
\end{tabular}
\end{table}

The local sweep also recovers the known context length and shows the
repeated-digit probe failing when its own output repeats the relevant
window. Resampling establishes detection at the hosted runs' sample
sizes for the stronger reference settings. Power is not universal: a
one-key watermark remains largely invisible even at 1 000 samples per
cell, whereas two keys suffice for detection in every trial at ten. The
flag threshold uses variation across prefixes and need not shrink with
sample size. Low entropy alone does not explain the field nulls: at nine
keys and 0.26 bits, below Fable's measured range, detection reaches 93\%
at ten samples. Full sweeps and power results are in
Appendix~\ref{app:C}.

\subsection{Redesigns expose native model behavior}\label{sec:3.2}

The described probe detects both local watermark variants but also
rejects on the clean model. Naming a digit in the instruction creates a
digit-to-fruit preference: variation in mean choice across digits is
twelve times larger than under echo wording. The statistic cannot
separate that native column structure from a watermark. The paper's more
conservative estimator stays clean but also loses detection on the
watermark variants (Appendix~\ref{app:C}).

The masking differential initially appears to repair this problem: it
detects both local watermark variants while remaining null on the clean
reference (Table~\ref{tab:1}). On Claude, however, it rejects on both
Fable and Haiku, including the \texttt{mixed} variant that never repeats
an entire number (Table~\ref{tab:2}). The control failure reveals a
confound that local validation did not capture.

\begin{table}[tbp]
\centering
\caption{Masking differential on Claude, fifty-two samples per
cell: same at \(H=12\), mixed at \(H=18\). Watermark labels reflect
provider statements. The Fable mixed run loses 12\% of requests to
refusals; the others have none. Here \(p\) is the 99\% upper bound from
5 000 permutations. The final column tests dependence of line one's
fruit on request type, before the intended repeated-context
comparison.}\label{tab:2}
\vspace{\baselineskip}
\begin{tabular}{@{}
  >{\raggedright\arraybackslash}p{(\linewidth - 10\tabcolsep) * \real{0.1622}}
  >{\raggedright\arraybackslash}p{(\linewidth - 10\tabcolsep) * \real{0.1216}}
  >{\raggedright\arraybackslash}p{(\linewidth - 10\tabcolsep) * \real{0.1351}}
  >{\raggedright\arraybackslash}p{(\linewidth - 10\tabcolsep) * \real{0.1081}}
  >{\raggedright\arraybackslash}p{(\linewidth - 10\tabcolsep) * \real{0.1081}}
  >{\raggedright\arraybackslash}p{(\linewidth - 10\tabcolsep) * \real{0.3649}}@{}}
\toprule
\begin{minipage}[b]{\linewidth}\raggedright
model
\end{minipage} & \begin{minipage}[b]{\linewidth}\raggedright
marked
\end{minipage} & \begin{minipage}[b]{\linewidth}\raggedright
variant
\end{minipage} & \begin{minipage}[b]{\linewidth}\raggedright
\(T\)
\end{minipage} & \begin{minipage}[b]{\linewidth}\raggedright
\(p\)
\end{minipage} & \begin{minipage}[b]{\linewidth}\raggedright
line-one fruit vs.~type, \(\chi^2\) (3 d.f.)
\end{minipage} \\
\midrule
Fable 5.1 & yes & same & 146.6 & 0.001 & 126.5 \\
Fable 5.1 & yes & mixed & 145.5 & 0.001 & 58.0 \\
Haiku 4.5 & no & same & 80.2 & 0.001 & 92.8 \\
Haiku 4.5 & no & mixed & 84.2 & 0.001 & 11.9 \\
\bottomrule
\end{tabular}
\end{table}

Neither Claude model repeats its first fruit on line two, and the first
choice already depends on the request type. Fable chooses bananas in
42\% of type A and 13\% of type B responses despite matched pooled
first-line digits. The instruction to write the same or different
numbers changes behavior before a context can repeat. The no-repetition
habit then transmits that difference to the measured choice.
Conditioning on the first fruit leaves a residual type effect on both
models, so this is not solely mediation through line one
(Appendix~\ref{app:D}). A prompt-identical design that lets the model
choose the first digit and classifies responses afterwards is a possible
repair, but was not tested. The transferable lesson is the need for a
control on the deployed model family.

\subsection{\texorpdfstring{Validation of the final
probe on a prompt-aware
reference}{Validation of the final probe on a prompt-aware reference}}\label{sec:3.3}

The deployment design of Section~\ref{sec:4}
combines the spaced probe, contexts that differ only in their last
digit, and the within-prefix score. At submission it had been validated
on the stock 3B reference only: echoed last-digit contexts detect stock
SynthID and stay at \(S = 1\) on the control and on the prompt-aware
variant, which masks echoed contexts by construction, and under the
spaced instruction the 3B model copies the number imperfectly, leaving
four to thirteen cells empty (Appendix~\ref{app:H}). We therefore
repeated the checks on Qwen3-8B-Instruct (4-bit) under the same keys and
\(h\), regenerating every response whose number differs from the request
(up to twelve of ninety at forty digits) and reading the choice
distribution exactly, so Table~\ref{tab:3} contains no sampling noise.

\begin{table}[tbp]
\centering
\caption{Design checks on Qwen3-8B-Instruct
(4-bit), peaches list, nine keys, \(h = 4\), contexts sharing every
digit but the last. Entries: the range of row-centred preferred-word
probabilities over contexts, from the exact distribution after one
verified copy per cell, and \(S\) under the code's estimator with the
published null at ten draws per cell (\(p_{99} = 0.0005\) for
\(S \geq 7\); \(S = 4\) at \(H = 8\), where the list is nearly
deterministic, gives 0.027).}\label{tab:3}
\vspace{\baselineskip}
\begin{tabular}{@{}
  >{\raggedright\arraybackslash}p{(\linewidth - 6\tabcolsep) * \real{0.1857}}
  >{\raggedright\arraybackslash}p{(\linewidth - 6\tabcolsep) * \real{0.2429}}
  >{\raggedright\arraybackslash}p{(\linewidth - 6\tabcolsep) * \real{0.2286}}
  >{\raggedright\arraybackslash}p{(\linewidth - 6\tabcolsep) * \real{0.3429}}@{}}
\toprule
\begin{minipage}[b]{\linewidth}\raggedright
probe, \(H\)
\end{minipage} & \begin{minipage}[b]{\linewidth}\raggedright
stock SynthID
\end{minipage} & \begin{minipage}[b]{\linewidth}\raggedright
control
\end{minipage} & \begin{minipage}[b]{\linewidth}\raggedright
SynthID, prompt-aware
\end{minipage} \\
\midrule
echo, 40 & 0.91, \(S = 10\) &
0.10, \(S = 0\) & 0.10,
\(S = 2\) \\
spaced, 4 & 0.79, \(S = 10\) &
0.36, \(S = 2\) & 0.79,
\(S = 9\) \\
spaced, 8 & 0.83, \(S = 4\) &
0.41, \(S = 1\) & 0.83,
\(S = 3\) \\
spaced, 40 & 0.86, \(S = 8\) &
0.59, \(S = 2\) & 0.86,
\(S = 9\) \\
\bottomrule
\end{tabular}
\end{table}

Two things follow. First, the positive control
exists: under the spaced instruction the prompt-aware variant is
indistinguishable from the stock variant at every length (column means
correlate at 1.00), whereas under the echo instruction it is
indistinguishable from the control. The watermark is active at the
choice whenever the model composes the number and masked whenever it
copies it. Second, the control is not clean. The unwatermarked model's
choice depends on the final digit in a direction shared across prefixes,
weakly under echo (0.10) and increasingly under the spaced wording as
the number grows (0.36, 0.41 and 0.59 at \(H = 4\), 8 and 40). At forty
digits the native range equals Fable's 0.57 (Section~\ref{sec:4.2}), and
at fifty draws per cell the within-prefix primary rejects on the
unwatermarked grid in five of five resamplings (\(S = 3\) to 6) while
the watermark variants keep \(S = 7\) to 9. On this reference the probe
separates watermark from no watermark by the size of \(S\) and of the
range, not by the \(p\)-value of the within-prefix null, and an
unwatermarked model can produce a forty-digit shift of the deployment's
magnitude.

\section{Deployment audit}\label{sec:4}

\subsection{Initial nulls and the final design}\label{sec:4.1}

The Haiku echo sweep over one to fifteen digits never rejects. Ten
initial Fable configurations vary context length and form, word list,
batch versus synchronous transport, and preceding output. None rejects;
one is uninformative because its choices collapse. Parsed responses
reproduce the context, and measured per-cell entropy is 0.6 to 0.95
bits. Appendix~\ref{app:D} gives every run and the explanations it
addresses. These nulls constrain only schemes covered by the probe's
assumptions; in particular, echo-based exclusions do not transfer to
prompt-aware masking.

The final design uses forty-digit contexts sharing their first
thirty-nine digits, presented in spaced form, and the within-prefix
score. It aims to preserve a long, newly constructed context while
minimising semantic differences among columns. Pilots showed that a
within-prefix chi-square also rejects on unwatermarked Claude and that
some lists fix half of Fable's prefixes on one word. A column-alignment
test and a within-prefix entropy pilot address those two problems. The
pilot selected
plums/\allowbreak{}limes/\allowbreak{}dates/\allowbreak{}kiwis at 0.99
bits per prefix (Appendix~\ref{app:H}).

We fixed a plan before collection, without external timestamping: fifty
samples per cell, a same-day Haiku control, the code's within-prefix
score with 10 000 permutations as primary, and a detection rule
requiring Fable \(p_{99}<0.01\), Haiku \(p_{99}>0.05\), nondegenerate
grids, and refusals not concentrated by context. Sonnet was added under
a later addendum. Batch over-requesting, retries, sample-size
differences and all other deviations are recorded in
Appendix~\ref{app:I}.

\subsection{A primary rejection with an inconclusive planned
decision}\label{sec:4.2}

\begin{figure}
\centering
\includegraphics[width=1\linewidth,height=\textheight,keepaspectratio,alt={The pre-registered run. Row-centred logit of the preferred word for every cell; outlined cells exceed the flag threshold. On Fable the flags fall in the columns for a final 3, 5 and 6 (red, below) and 2 (green, above), in five of the ten prefixes; on Haiku no column stands out.}]{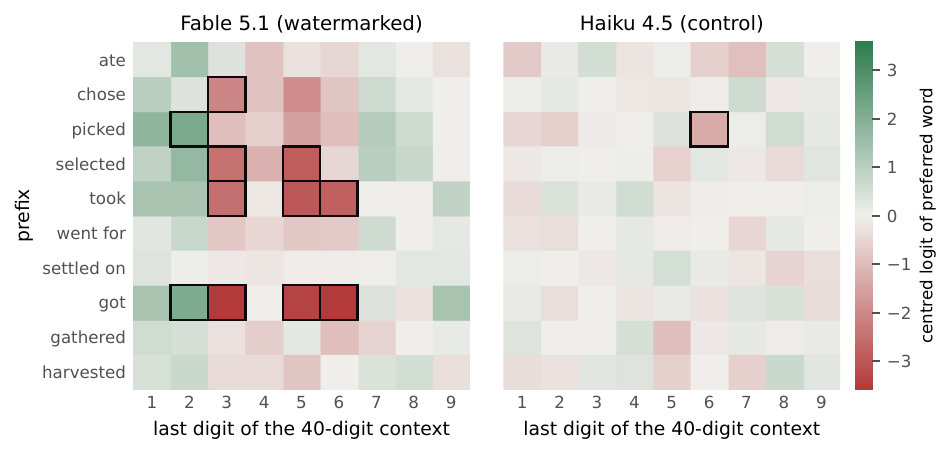}
\caption{The pre-registered run. Row-centred logit of the preferred word
for every cell; outlined cells exceed the flag threshold. On Fable the
flags fall in the columns for a final 3, 5 and 6 (red, below) and 2
(green, above), in five of the ten prefixes; on Haiku no column stands
out.}\label{fig:1}
\end{figure}

Table~\ref{tab:4} reports the final grids. The primary rejects on Fable
at its permutation bound's floor, while both controls remain null.
Figure~\ref{fig:1} shows the column alignment. The effect depends on the
estimator: the same score with the paper's \(\hat\sigma\) gives
\(p_{99}=0.055\). Permutation increases the code's estimator and reduces
flags, whereas it leaves the paper's estimator unchanged. The primary's
\(p\)-value thus belongs to the entire procedure rather than the
observed flag count alone. Because \(S\) counts
whole cells, at \(S = 4\) one or two cells separate the two
\(p\)-values; the continuous effect size below does not share this
instability.

\begin{table}[tbp]
\centering
\caption{Final spaced-context audit with
plums/\allowbreak{}limes/\allowbreak{}dates/\allowbreak{}kiwis. Counts
are received responses; parsing succeeds for 4 965 Fable, 5 845 Haiku
and 2 300 Sonnet responses. Fable planned fifty per cell and received 1
to 65 after refusals and retries; Haiku received sixty-five. Sonnet used
eight well-sampled prefixes under an addendum, planning twenty-five per
cell and receiving thirty-two. The primary is the within-prefix score
with the code estimator; the last column changes only the estimator.
Full scores and deviations are in Appendices \ref{app:E} and
\ref{app:I}.}\label{tab:4}
\vspace{\baselineskip}
\begin{tabular}{@{}
  >{\raggedright\arraybackslash}p{(\linewidth - 8\tabcolsep) * \real{0.1310}}
  >{\raggedright\arraybackslash}p{(\linewidth - 8\tabcolsep) * \real{0.1429}}
  >{\raggedright\arraybackslash}p{(\linewidth - 8\tabcolsep) * \real{0.1190}}
  >{\raggedright\arraybackslash}p{(\linewidth - 8\tabcolsep) * \real{0.2857}}
  >{\raggedright\arraybackslash}p{(\linewidth - 8\tabcolsep) * \real{0.3214}}@{}}
\toprule
\begin{minipage}[b]{\linewidth}\raggedright
model
\end{minipage} & \begin{minipage}[b]{\linewidth}\raggedright
responses
\end{minipage} & \begin{minipage}[b]{\linewidth}\raggedright
refused
\end{minipage} & \begin{minipage}[b]{\linewidth}\raggedright
primary \(S\), \(p_{99}\)
\end{minipage} & \begin{minipage}[b]{\linewidth}\raggedright
paper estimator \(S\), \(p_{99}\)
\end{minipage} \\
\midrule
Fable & 4 965 & 44\% & \textbf{\boldmath 4, 0.0005} & 4, 0.055 \\
Haiku & 5 850 & 0 & 1, 1.00 & 3, 0.52 \\
Sonnet 5 & 2 304 & 0 & 1, 1.00 & 1, 1.00 \\
\bottomrule
\end{tabular}
\end{table}

The planned decision is nevertheless \textbf{\boldmath inconclusive}. Fable's
first-round refusals are concentrated by context (78 to 177 by final
digit, chi-square \(p<10^{-10}\)), violating the refusal criterion. A
filter that preferentially removes particular responses could create
column structure. Refusals return no text; their apparent dependence on
the prompt, inferred from the echo comparison below, does not rule out
selection bias. Refusal rate is uncorrelated with row-centred share
(\(r=-0.03\) over ninety cells), and low-refusal cells retain the
pattern, but these checks are exploratory and cannot restore the planned
verdict.

As a continuous effect size, we use the maximum minus minimum column
mean of the row-centred share of each model's pooled most frequent word,
on eight shared prefixes. It is 0.57 on Fable (percentile bootstrap 0.53
to 0.61, biased upward for a range), versus 0.07 on Haiku and 0.12 on
Sonnet. Within-prefix permutation gives \(p_{99}=0.0005\), 0.78 and
0.20, respectively. Restricting to twenty-two first-round verbatim
responses per cell on eight well-sampled prefixes preserves a range of
0.50 to 0.66 over one hundred subsamples; the primary rejects at 0.01 in
51 of those draws. The effect survives this check, while the discrete
decision is unstable at the reduced size. Full estimates, split-half
comparisons and exclusion checks are in Appendix~\ref{app:E}.

The controls also differ in instruction following: Fable reproduces the
number verbatim in 99.7\% of responses, whereas 55\% of Haiku and 20\%
of Sonnet responses are nonverbatim. Restricted to exact copies, the
controls' primary \(p_{99}\) values are 1.00 and 0.08. Both models show
within-prefix digit dependence under chi-square, reinforcing the
distinction between a native digit preference and cross-prefix
alignment. Thus the results support a substantial
context-dependent shift on Fable that was absent on two siblings while
they were unmarked; Section~\ref{sec:4.4} re-queries them after marking
and Section~\ref{sec:4.5} weighs the explanations.

\subsection{What follow-ups establish}\label{sec:4.3}

Echoing the same compact number from the prompt (E40) gives no refusals
and no rejection, consistent with prompt-aware masking. The comparison
also changes number salience and filtering, so it does not isolate
masking. Changing the base number (W40c) gives an inconclusive
replication under the plan: its primary does not reject and refusals
remain concentrated, although its exploratory range rejects with a
different column pattern. Varying a digit triple inside or outside a
proposed context window gives two further primary nulls. Their
continuous scores are confounded by conspicuous strings, a behavior also
observed on an unwatermarked control (Appendices \ref{app:E} and
\ref{app:H}).

\begin{table}[tbp]
\centering
\caption{Follow-up grids on Fable: eight prefixes, twenty-five
samples per cell, the list of Table~\ref{tab:4}; the inside and outside
grids hold the last digit at 4. Entries are \(S\), \(p_{99}\) under the
primary and under the paper's estimator, and the continuous range with
its within-prefix permutation \(p_{99}\). E40 is null; the other three
are inconclusive under the plan, and both varied-triple grids fail its
refusal condition. Appendix~\ref{app:H} gives additional design details
and decision rules.}\label{tab:5}
\vspace{\baselineskip}
\begin{tabular}{@{}
  >{\raggedright\arraybackslash}p{(\linewidth - 10\tabcolsep) * \real{0.1053}}
  >{\raggedright\arraybackslash}p{(\linewidth - 10\tabcolsep) * \real{0.2842}}
  >{\raggedright\arraybackslash}p{(\linewidth - 10\tabcolsep) * \real{0.1053}}
  >{\raggedright\arraybackslash}p{(\linewidth - 10\tabcolsep) * \real{0.1053}}
  >{\raggedright\arraybackslash}p{(\linewidth - 10\tabcolsep) * \real{0.2421}}
  >{\raggedright\arraybackslash}p{(\linewidth - 10\tabcolsep) * \real{0.1579}}@{}}
\toprule
\begin{minipage}[b]{\linewidth}\raggedright
grid
\end{minipage} & \begin{minipage}[b]{\linewidth}\raggedright
change from Table~\ref{tab:4}
\end{minipage} & \begin{minipage}[b]{\linewidth}\raggedright
refused
\end{minipage} & \begin{minipage}[b]{\linewidth}\raggedright
primary
\end{minipage} & \begin{minipage}[b]{\linewidth}\raggedright
paper's \(\hat\sigma\)
\end{minipage} & \begin{minipage}[b]{\linewidth}\raggedright
range
\end{minipage} \\
\midrule
E40 & compact number spelled in the prompt & 0 & 1, 1.00 & 2, 0.75 &
0.10, 0.48 \\
W40c & second base number & 37\% & 1, 1.00 & 4, 0.011 & 0.51, 0.0005 \\
inside & triple varied at digits 28--30 & 20\% & 2, 0.48 & 2, 0.85 &
0.55, 0.001 \\
outside & triple varied at digits 1--3 & 9\% & 0, 1.00 & 2, 0.49 & 0.22,
0.004 \\
\bottomrule
\end{tabular}
\end{table}

A length search yields only a conditional upper bound of fourteen
tokens, assuming that the shift is a context hash and that the inferred
segmentation and measured slot are correct. Selective refusals, changing
hash inputs and an unmet stopping rule prevent a lower bound or
identification of the window (Appendix~\ref{app:E}).

\begin{table}[tbp]
\centering
\caption{Bisection over the digit length on Fable. Each entry
is \(S\), \(p_{99}\); ``code'' and ``paper'' are the two \(\hat\sigma\)
estimators, the code's within-prefix score being the pre-registered
primary. \(D = 25\), \(28\) and \(31\) were sampled twice and pooled;
the \(D = 40\) row is the grid of Table~\ref{tab:4} restricted to the
same eight prefixes. Windows count the two space tokens. At \(D = 28\)
one cell is empty and refusals are concentrated by context; at
\(D = 31\) three contexts are depleted.}\label{tab:6}
\vspace{\baselineskip}
\begin{tabular}{@{}
  >{\raggedright\arraybackslash}p{(\linewidth - 10\tabcolsep) * \real{0.0769}}
  >{\raggedright\arraybackslash}p{(\linewidth - 10\tabcolsep) * \real{0.0989}}
  >{\raggedright\arraybackslash}p{(\linewidth - 10\tabcolsep) * \real{0.1319}}
  >{\raggedright\arraybackslash}p{(\linewidth - 10\tabcolsep) * \real{0.1099}}
  >{\raggedright\arraybackslash}p{(\linewidth - 10\tabcolsep) * \real{0.2967}}
  >{\raggedright\arraybackslash}p{(\linewidth - 10\tabcolsep) * \real{0.2857}}@{}}
\toprule
\begin{minipage}[b]{\linewidth}\raggedright
\(D\)
\end{minipage} & \begin{minipage}[b]{\linewidth}\raggedright
window
\end{minipage} & \begin{minipage}[b]{\linewidth}\raggedright
responses
\end{minipage} & \begin{minipage}[b]{\linewidth}\raggedright
refused
\end{minipage} & \begin{minipage}[b]{\linewidth}\raggedright
within-prefix, code / paper
\end{minipage} & \begin{minipage}[b]{\linewidth}\raggedright
published, code / paper
\end{minipage} \\
\midrule
25 & 11 & 4 602 & 0.1\% & 3, 0.059 / 4, 0.017 & 3, 0.10 / 4, 0.0041 \\
28 & 12 & 3 704 & 48\% & 1, 0.99 / 4, 0.069 & 1, 1.00 / 4, 0.029 \\
31 & 13 & 3 684 & 53\% & 0, 1.00 / 3, 0.38 & 0, 1.00 / 3, 0.23 \\
34 & 14 & 2 040 & 34\% & \textbf{\boldmath 4, 0.0007} / 4, 0.0084 & 5, 0.0007 / 5,
0.0007 \\
40 & 16 & 4 609 & 44\% & \textbf{\boldmath 4, 0.0005} / 3, 0.22 & 4, 0.0056 / 3,
0.021 \\
\bottomrule
\end{tabular}
\end{table}

At submission the combination of the spaced probe
and within-prefix score had \textbf{\boldmath no positive control on a
prompt-aware reference}: the small local model follows the spaced
instruction imperfectly, and its scorer did not verify the number.
Section~\ref{sec:3.3} supplies that control on a larger reference with
the number verified, and shows at the same time that an unwatermarked
reference has native column structure under this probe, which
Section~\ref{sec:4.5} takes into account.

\subsection{\texorpdfstring{The former controls
after marking}{The former controls after marking}}\label{sec:4.4}

By November the provider's support table listed text
watermarks for Haiku 4.5 and Sonnet 5 as well (Appendix~\ref{app:A.3}).
On 8 November 2026 we re-queried both identifiers with the design of
Table~\ref{tab:4} (same number, list and sample sizes) under a plan
fixed beforehand (\texttt{prereg\_W40b\_recheck.md};
Appendix~\ref{app:E.2}), and added Sonnet 5.5 and Haiku 5.5, marked at
launch, as exploratory grids. No request was refused.

\begin{table}[tbp]
\centering
\caption{The design of Table~\ref{tab:4} on
the former controls before and after the provider listed them as marked,
and on two models marked at launch; 2 304 responses per Sonnet grid
(eight prefixes), 5 850 per Haiku 4.5 grid, 6 955 for Haiku 5.5 (4 774
parsed). Range and within-prefix \(p_{99}\) as in Section~\ref{sec:4.2}
for the grid's most frequent word; \(r\) correlates the nine kiwis
column means with Fable's (Table~\ref{tab:4} grid).}\label{tab:7}
\vspace{\baselineskip}
\begin{tabular}{@{}
  >{\raggedright\arraybackslash}p{(\linewidth - 10\tabcolsep) * \real{0.1398}}
  >{\raggedright\arraybackslash}p{(\linewidth - 10\tabcolsep) * \real{0.1290}}
  >{\raggedright\arraybackslash}p{(\linewidth - 10\tabcolsep) * \real{0.0968}}
  >{\raggedright\arraybackslash}p{(\linewidth - 10\tabcolsep) * \real{0.2581}}
  >{\raggedright\arraybackslash}p{(\linewidth - 10\tabcolsep) * \real{0.1935}}
  >{\raggedright\arraybackslash}p{(\linewidth - 10\tabcolsep) * \real{0.1828}}@{}}
\toprule
\begin{minipage}[b]{\linewidth}\raggedright
model
\end{minipage} & \begin{minipage}[b]{\linewidth}\raggedright
run
\end{minipage} & \begin{minipage}[b]{\linewidth}\raggedright
marked
\end{minipage} & \begin{minipage}[b]{\linewidth}\raggedright
primary \(S\), \(p_{99}\)
\end{minipage} & \begin{minipage}[b]{\linewidth}\raggedright
range, \(p_{99}\)
\end{minipage} & \begin{minipage}[b]{\linewidth}\raggedright
\(r\) with Fable
\end{minipage} \\
\midrule
Sonnet 5 & September &
no & 1, 1.00 &
0.12, 0.20 & \(+0.53\) \\
Sonnet 5 & November &
yes & 4, 0.016 &
0.47, 0.0005 & \(+0.86\) \\
Haiku 4.5 & September &
no & 1, 1.00 &
0.07, 0.78 & \(-0.11\) \\
Haiku 4.5 & November &
yes & 0, 1.00 &
0.13, 0.023 & \(+0.52\) \\
Sonnet 5.5 & November &
yes & 5, 0.0005 &
0.48, 0.0005 & \(+0.78\) \\
Haiku 5.5 & November &
yes & 0, 1.00 &
0.39, 0.0005 & \(+0.89\) \\
\bottomrule
\end{tabular}
\end{table}

Sonnet 5 changed on identical prompts: the primary
moved from \(S = 1\), \(p_{99} = 1.00\) to \(S = 4\), \(p_{99} = 0.016\)
(on verbatim copies \(S = 6\), \(p_{99} = 0.0015\)), the range from 0.12
to 0.47, and the column means of the kiwis share now follow Fable's at
\(r = +0.86\) (limes: \(+0.97\)), with the same preferred word. Under
the plan's rules this grid is inconclusive, the primary being in the
grey zone, while the continuous statistic shows a shift of Fable's size.
Sonnet 5.5 meets the plan's detection rule and shares the pattern
(\(r = +0.78\) with Fable, \(+0.76\) with Sonnet 5). Haiku 4.5 meets its
``shift absent'' rule (\(S = 0\); kiwis range 0.07 before and after, so
its \(r\) is a correlation of noise). Haiku 5.5 prefers kiwis strongly
(top share 0.76), so its discrete statistic stays at \(S = 0\), yet the
continuous statistic shows the same pattern (range 0.39, \(r = +0.89\);
on verbatim copies the primary gives \(S = 2\), \(p_{99} = 0.001\)). The
token-counting endpoint segments the forty digits identically on all
five models (increments at \(k = 1, 4, 7, \dots, 40\)), so shared keys
would give shared colours. The re-check cannot show that any response
carries a mark in the provider's sense; it shows that the probe's
pattern appears on an unchanged identifier when the provider marks it
and on three further marked models, while one marked model stays
silent.

\subsection{\texorpdfstring{Three explanations and
what separates
them}{Three explanations and what separates them}}\label{sec:4.5}

Three accounts fit the September evidence, and the
re-check bears on each. \emph{A context-keyed watermark with
prompt-blind history on this route.} For: the shift aligns across
prefixes, recurs at \(D = 34\) and in split halves, and was absent on
two siblings while they were unmarked; the reference shows that the
spaced probe detects exactly this signature, under prompt-aware masking
included (Section~\ref{sec:3.3}); and when the provider marked Sonnet 5,
the same identifier, on the same prompts and without a refusal, acquired
a shift of Fable's size whose column pattern matches Fable's, Sonnet
5.5's and Haiku 5.5's, as shared keys and shared tokenization predict.
Against: Haiku 4.5, also listed as marked, shows nothing; and the
described pilots with twenty-four and forty copies of one digit compose
their contexts and were null, though at sample sizes with power near one
half (Appendix~\ref{app:H}). \emph{A native preference of Fable for the
final digit.} For: an unwatermarked reference produces a forty-digit
range of 0.59 under this very probe and loses it under echo wording
(Table~\ref{tab:3}), so neither the size of the shift nor the E40 null
separates this account from the first. Against: Sonnet 5's pattern
appeared between September and November on an unchanged identifier and
matches Fable's, which a preference cannot do; the unmarked siblings'
digit preferences did not align across prefixes; and the second base
number changed the column pattern (\(r = -0.34\)), as a hash predicts.
\emph{Selection by the refusal filter.} For: Fable's refusals are
concentrated by context and the filter's inputs are unknown. Against:
the checks of Section~\ref{sec:4.2}, and the same pattern on Sonnet 5
and Sonnet 5.5 with no refusals.

Open: whether Fable's responses carry a mark, which
only the provider's detector can say; why marked Haiku 4.5 is null; and
whether a prompt-identical two-line design (Section~\ref{sec:3.2})
reproduces the shift. The evidence now favours the first account; the
pre-registered decision for Fable stays inconclusive because its refusal
condition failed.

\section{Interpretation and limitations}\label{sec:5}

\subsection{A checklist for presence-test audits}\label{sec:5.1}

The experiments motivate eight checks before treating a null or a
rejection as evidence about a deployment:

\begin{enumerate}
\def\labelenumi{\arabic{enumi}.}
\tightlist
\item
  \textbf{\boldmath Validate the exact protocol.} Establish detection on a
  reference scheme at the measured position and sample size;
  component-level validation does not cover an untested combination.
\item
  \textbf{\boldmath Use same-family controls.} A clean surrogate cannot exclude
  native preferences or response-planning confounds on the deployed
  model.
\item
  \textbf{\boldmath Check uncertainty within cells.} Pooled word counts can hide
  deterministic choices that the watermark cannot move.
\item
  \textbf{\boldmath Resolve context units.} Digit counts, token boundaries and
  constant spaces change the controlled window.
\item
  \textbf{\boldmath State deployment assumptions.} Key stability, context reach
  and prompt-history handling determine what a null can exclude; each
  unsupported assumption limits the claim.
\item
  \textbf{\boldmath Audit missingness.} Context-dependent refusals and format
  failures can affect validity as well as power.
\item
  \textbf{\boldmath Check differential conditions.} A prompt that reveals the
  comparison type can change behavior before the position intended to
  isolate the watermark.
\item
  \textbf{\boldmath Inspect the effect across contexts.} A conspicuous single
  string can dominate a continuous statistic; report the pattern and
  controls alongside its significance.
\end{enumerate}

Prompt-aware masking alone does not explain every deployment null. Long
described-context pilots also fail, at small sample sizes with
low-entropy lists. The proposed account additionally requires adequate
context length and a score that tolerates prefix effects; these
observations do not identify Claude's implementation. If the final shift
were attributed to a watermark, it would exclude an inactive route and a
fresh independent key per request, but not a small key pool. Such
conclusions remain conditional on attribution.

\subsection{Scope and evidential limits}\label{sec:5.2}

The study covers one stated-watermarked model over
one week in September and a one-day re-check of four models in
November. Provider statements determine the controls' assumed status;
the detection API was unavailable, so no hosted response has keyed
ground truth. Reference power checks cannot exclude a deployment with a
weaker bias. The primary effect rests on one planned
grid per model, with an inconclusive decision; W40c and the bisection
grids replicate the continuous shift but not the primary decision.
Unequal cell sizes and selective refusals remain material limitations on
Fable; the reference shows native column structure under the final probe
(Section~\ref{sec:3.3}); and the re-check rests on the provider's
statement that the re-queried identifiers are marked and otherwise
unchanged, with marked Haiku 4.5 unexplained.

Only the Red-Green test was implemented; Fixed-Sampling and
Cache-Augmented presence tests, ordinary-word contexts, and the
prompt-identical differential remain untested. A provider detector could
establish whether the responses carry a watermark; it would not by
itself establish that watermarking caused the observed shift.

\section{Conclusion}\label{sec:6}

A presence test is a statistical argument whose
premises a deployment can falsify. Seeding the repeated-context history
from the prompt leaves the watermark active and blinds the published
probe, so an echo-based null excludes only prompt-blind schemes. Two
redesigns correct on every reference variant fail on the deployed
family. The pre-registered audit then finds a large, internally
consistent, context-dependent shift on Fable that two unmarked siblings
did not show; the primary rejects, but refusals concentrated by context
void the planned decision, reported as inconclusive. After the provider
marked the siblings, Sonnet 5 acquired, without refusals, a shift of
Fable's size and column pattern, as did two models marked at launch;
Haiku 4.5 did not. We read the Fable shift as most plausibly a
context-keyed watermark with shared keys, noting that an unwatermarked
reference produces a shift of this size under the same probe, so
same-model contrasts, not same-family controls, settle attribution.
Presence testing is informative only about schemes covered by explicit
assumptions and controls; we have requested detector access to test mark
presence directly.

\FloatBarrier

\label{main-text-end} \clearpage

\subsection*{AI use statement}

Generative AI tools were used for language editing,
manuscript review, restructuring and shortening of the presentation,
preparation of the LaTeX manuscript, build scripts and submission
artifacts, and, for the revision, for the scripts that collected and
scored the follow-up runs of Sections 3.3 and 4.4, which the authors
specified and verified. They were not used to design the study or to
develop its original research code. Numerical statements were checked
against the stored responses, tables and analysis code. The authors take
responsibility for the final text, claims and artifacts.

\subsection*{Ethics statement}

Queries used ordinary API access within the provider's terms, without
attempts to recover keys, spoof or remove a mark. Presence testing can
also inform watermark-stealing attacks; this study reports test
limitations and native-behavior confounds, and distinguishes statistical
rejection from attribution to a watermark.

\subsection*{Reproducibility statement}

Appendix~\ref{app:A} specifies the statistics and implementation;
Appendix~\ref{app:B} documents the tokenizer measurements; and
Appendix~\ref{app:F} provides the probe prompts and word lists.
Appendices C to E, G and H preserve the detailed results and robustness
checks. Appendix~\ref{app:I} identifies the scoring code, stored
response matrices, pre-registration plans and collection deviations. The
anonymous supplementary archive includes these files,
\texttt{analysis/equal\_n.py}, an offline reproduction guide and a
SHA-256 manifest. Statistics can be recomputed without new API calls.
The November re-check and the Qwen3-8B checks (runs
\texttt{*\_recheck*} and \texttt{qwen8b/}, plan
\texttt{prereg\_W40b\_recheck.md}, \texttt{analysis/recheck.py} and
\texttt{analysis/local\_probe\_stats.py}) postdate the archive and are
in the repository. The public repository link is omitted for anonymous
review.

\clearpage

\phantomsection\label{references-start}

\bibliographystyle{iclr2027_conference}
\bibliography{references}
\clearpage
\appendix

\section{Statistical definitions and implementation}\label{app:A}

\subsection{SynthID-Text}\label{app:A.1}

SynthID-Text \citep{dathathri2024} modifies the model's next-token
distribution \(p_t\) using \(m\) secret keys. For each candidate \(x\)
and context \(c_t = (y_{t-h}, \dots, y_{t-1})\) of \(h\) preceding
tokens it computes

\[g_\ell (x, c_t) = \mathrm{hash} (k_\ell, c_t, x) \bmod 2, \qquad \ell = 1, \dots, m.\]

A tournament over \(2^m\) candidates drawn from \(p_t\) favours
\(g_\ell = 1\) in round \(\ell\), breaking ties randomly. The winner's
distribution therefore has a bias fixed by the keys and context. A keyed
detector averages \(g_\ell (y_t,c_t)\) over the text; the mean is
\(1/2\) without watermarking and higher with it.

The second property relevant here is \emph{repeated-context masking}:
after a context has occurred in the generation, its next occurrence is
sampled from \(p_t\) unchanged. Marking each context at most once
preserves the text distribution in expectation over the key
\citep[Supplementary Information G]{dathathri2024}, and lets a probe
suppress marking at a position by repeating its context.

What counts as a previous occurrence depends on the implementation. The
\texttt{transformers} reference starts with an empty rolling window and
context-hash history; neither is seeded from the prompt. A context
copied from the prompt thus counts as new. Section~\ref{sec:3.1} changes
this one property.

\subsection{The Red-Green statistic}\label{app:A.2}

The Red-Green presence test of \citet{gloaguen2025} uses the first
property mentioned above. During the test evaluation the LLM is asked to
choose a word from a set \(\Sigma\) after a sentence that the tester
controls. The last few tokens of the sentence are the \emph{context}
\(t_2\), and its earlier part is the \emph{prefix} \(t_1\). Let
\(p_{t_1,t_2} (x)\) be the probability of choosing \(x \in \Sigma\). The
authors model its logit as

\[\ell_{t_1,t_2} (x) = \frac{x^0}{\tau} + \delta'_{t_2} (x) + \varepsilon'_{t_1,t_2} (x),\]

Here \(x^0\) is the model's own preference for \(x\), \(\tau\) is the
temperature, and \(\varepsilon'\) is independent noise with variance
\(\sigma^2\). The term \(\delta'_{t_2} (x)\) is the shift of the
watermarking algorithm that depends on the context alone, and is equal
to zero if there is no watermark. For a Red-Green scheme with a fixed
boost \(\delta\) the shift takes values in
\(\{-\delta/\tau, 0, +\delta/\tau\}\), according to the colors of \(x\)
and its competitors. For SynthID the shift varies with the tournament,
but its sign under a given context is still fixed.

The probe is one instruction:

\begin{quote}
Complete the sentence ``\(t_1\) \(t_2\)'' using a random word from:
{[}\(\Sigma\){]},
\end{quote}

\begin{figure}
\centering
\includegraphics[width=1\linewidth,height=\textheight,keepaspectratio,alt={One probe response on the two tokenizers used in this paper. The watermark biases the choice as a function of the h tokens before it; the probe arranges for those tokens to be the controlled context t\_2, so that the bias varies by column and not by row. On Claude the window includes a constant space token and the digits arrive three to a token.}]{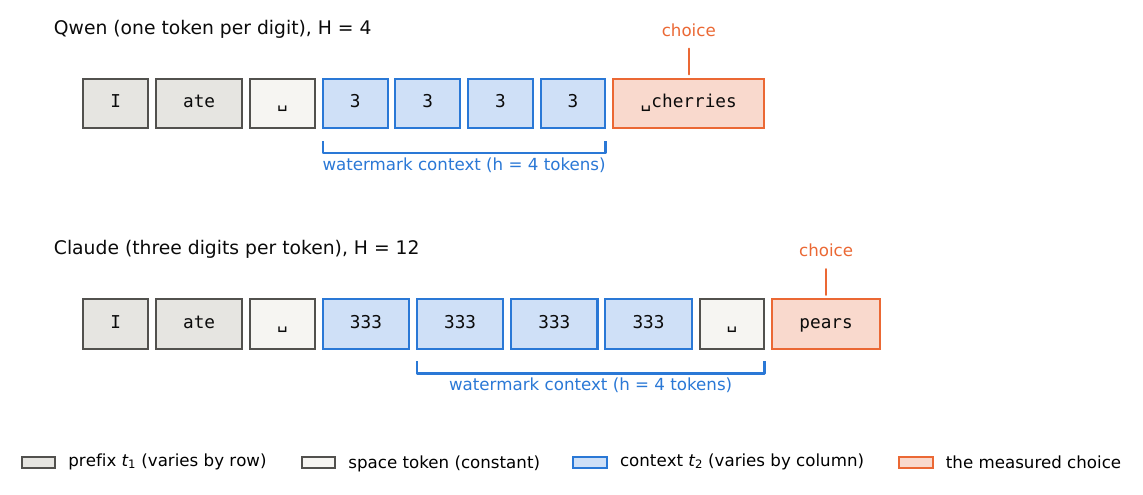}
\caption{One probe response on the two tokenizers used in this paper.
The watermark biases the choice as a function of the \(h\) tokens before
it; the probe arranges for those tokens to be the controlled context
\(t_2\), so that the bias varies by column and not by row. On Claude the
window includes a constant space token and the digits arrive three to a
token.}\label{fig:2}
\end{figure}

In the baseline version \(t_2\) is one of \(N_2 = 9\) digits repeated
\(H\) times, \(t_1\) is one of \(N_1 = 10\) verb phrases like ``I ate'',
and the set \(\Sigma\) has four words. The model echoes the sentence and
appends its choice, so the chosen word follows \(t_2\) directly. Each
pair \((t_1, t_2)\) is one prompt, which we call a \emph{cell}, and each
prompt is sampled \(N\) times. The preferred word \(x^\ast\) is the word
with the largest logit summed over all cells. Its empirical probability
in each cell is converted to a logit. The \(N_1 \times N_2 = 90\) logits
form a matrix \(L \in \mathbb{R}^{N_1 \times N_2}\): its rows are
prefixes and its columns are contexts. Figure~\ref{fig:2} shows one
response as the watermark sees it.

The statistic looks for column structure in \(L\): each row is centred
on its own median, which removes \(x^0\) and any prefix-specific taste.
Then the noise level \(\hat\sigma\) is estimated from the matrix itself.
The paper takes the square root of the median over \(t_1\) of the
variance across \(t_2\) (their Eq. 5); permuting contexts within a row
leaves it unchanged. The authors' released code takes the median over
\(t_2\) of the standard deviation across \(t_1\), and we report both.
The second is smaller under a watermark, so it flags more cells, and,
being the standard deviation of uncentred logits across prefixes,
includes the prefix effect; the first is usually more robust to genuine
column structure in the model, though not always (Table~\ref{tab:18}).
Throughout, \(p_{99}\) denotes the 99\% Clopper-Pearson upper bound of
the permutation \(p\)-value. Cells below \(-r\hat\sigma\) are flagged
red and cells above \(+r\hat\sigma\) green, with \(r = 1.96\). Let
\(\mathrm{red} (t_2)\) and \(\mathrm{green} (t_2)\) be the numbers of
red and green cells in column \(t_2\). The statistic, as the released
code computes it, is

\[S (L) = \max\bigl (\max_{t_2} \mathrm{red} (t_2),\ \max_{t_2} \mathrm{green} (t_2)\bigr) - \max\bigl (\min_{t_2} \mathrm{red} (t_2),\ \min_{t_2} \mathrm{green} (t_2)\bigr).\]

Under a watermark some context colors \(x^\ast\) red for every prefix
while some other context colors it red for none, so \(S\) approaches
\(N_1\). Under the null the flags are the tails of the noise, and \(S\)
stays small. Under the null of the model above the cells are
exchangeable, so the \(p\)-value comes from permuting the cells of \(L\)
and recomputing \(S\); on a model with strong prefix effects they are
not, which Section~\ref{sec:4} addresses. We report the upper end of a
99\% Clopper-Pearson interval on the proportion of permutations that
reach the observed value. The test rejects only when the flagged cells
line up with \(t_2\) in a way that random rearrangement does not
reproduce.

The test is correct under its model, and the model has assumptions. Six
of them recur in this paper.

\begin{enumerate}
\def\labelenumi{\arabic{enumi}.}
\tightlist
\item
  \emph{Entropy.} The choice must be genuinely uncertain: a
  near-deterministic choice gives the watermark nothing to shift, and
  its small fluctuations look large against a small \(\hat\sigma\).
\item
  \emph{Containment.} The watermark's window of \(h\) tokens must lie
  inside \(t_2\). If it reaches into \(t_1\), the shift depends on the
  prefix and the column structure disappears.
\item
  \emph{Exact length.} For SynthID, \(H\) must equal \(h\). A longer run
  makes the window before the choice a repeat of the window one token
  earlier. Masking then switches the watermark off exactly where the
  probe looks \citep[App. F.4]{gloaguen2025}.
\item
  \emph{One key.} The \(N\) samples in a cell and the \(N_1\) cells in a
  column must be produced under the same key, or under a small pool of
  keys whose average bias is stable \citep[App. F.2]{gloaguen2025}.
\item
  \emph{No native column structure.} The unwatermarked model must not
  itself prefer particular words after particular contexts. Such a
  preference cannot be told apart from a watermark.
\item
  \emph{An active watermark at the probe position.} The shift must be
  applied at the step the probe measures.
\end{enumerate}

The original authors discuss the first five, or they follow from their
model. The sixth is implicit and it is the one a deployment can violate.

\subsection{Provider disclosures}\label{app:A.3}

Anthropic announced default SynthID-Text-based marking with no opt-out
for models released after 2 August 2026, described as surviving light
edits but not a complete rewrite \citep{nemecek2026}. Its announcement,
updated on 1
September,\footnote{Anthropic, \emph{How Claude's text watermark works}: \url{https://www.anthropic.com/news/claude-text-watermark}.}
placed the detection API in a private preview for eligible
organisations, which did not include us, and said older models would
receive marking over the coming months. None was stated to have received
it during our study; the Models API then listed one model after the
cutoff, \texttt{claude-fable-5-1}, created on 28 August 2026. As of 23
September 2026, the provider's support table lists text watermarks for
Fable 5.1 but not for Haiku 4.5 or Sonnet 5 on first-party surfaces,
including the
API.\footnote{Anthropic, \emph{How Claude marks AI-generated content}, accessed 23 September 2026: \url{https://support.claude.com/en/articles/16266773-how-claude-marks-ai-generated-content}.}
It plans to cover all pre-cutoff models by 2 December 2026. This
supports the controls' assumed status as of the access date; it does not
guarantee that they will remain unmarked until December.
Read again on 8 November 2026, the same table lists
text watermarks for Haiku 4.5 and Sonnet 5 on first-party surfaces,
including the API, and for Sonnet 5.5 and Haiku 5.5 from their release;
it does not state when marking was enabled for the older models or
whether anything else about their serving changed. Section~\ref{sec:4.4}
uses this change.

The provider discloses neither context length or units, key count or
schedule, repeated-context handling, nor conditions for skipping
marking. The test's corresponding assumptions cannot be verified
directly.

\subsection{Endpoint contract}\label{app:A.4}

The local reference is Qwen2.5-3B-Instruct with SynthID-Text, where
ground truth is known. The hosted models are \texttt{claude-fable-5-1}
(Fable), stated to be watermarked; \texttt{claude-haiku-4-5} (Haiku),
released in October 2025; and \texttt{claude-sonnet-5} (Sonnet),
released on 29 June 2026 and added as a second control in
Section~\ref{sec:4}. The controls' assumed status follows the cutoff in
Appendix~\ref{app:A.3}. Hosted runs span 6 to 13
September 2026; the re-check of Section~\ref{sec:4.4} ran on 8 November
2026 on \texttt{claude-haiku-4-5}, \texttt{claude-sonnet-5},
\texttt{claude-sonnet-5-5} and \texttt{claude-haiku-5-5}.

The test requires independent samples for each prompt. We follow the
authors' released code, also report the paper's \(\hat\sigma\)
estimator, use a Clopper-Pearson upper bound for permutation
\(p\)-values, and add the safeguards of Section~\ref{sec:2.3}. Locally,
we generate one response per prompt, locate the choice after its last
digit-token run, renormalise the processor's distribution over the four
words' first tokens, and draw \(N\) samples. This simulates estimator
noise conditional on that generated response. The slot finder does not
verify the requested number: echo responses reproduce it, but this
limits validation of the spaced form (Appendix~\ref{app:H}).
The Qwen3-8B checks of Section~\ref{sec:3.3} add
that verification. The same weights supply the stock watermark, clean
control and prompt-aware variant.

Hosted completions are sampled through the Anthropic API, synchronously
or in batches. Sampling parameters cannot be set on this Claude
generation. Fable's extended thinking cannot be disabled; at the lowest
effort, usage counts observed during collection matched the visible
answer length, indicating no reasoning tokens. Usage was not logged. We
do not request the API's opt-in prompt caching, and a nonce at the
system prompt's start prevents requests sharing a cached prefix while
staying outside the measured context. Fable's safety classifier refuses
0.1\% to 96\% of requests depending on the prompt family and digit
string (Appendix~\ref{app:G}). Refusals are dropped without a fallback
model.

\subsection{Probe definitions}\label{app:A.5}

All four probes measure a choice from a four-word list after a
controlled context (full prompts in Appendix~\ref{app:F}).

\textbf{\boldmath Echo.} The published probe spells out the sentence to reproduce.
The context occurs in both the instruction and the format example; only
the example includes the trailing space of the measured window.

\textbf{\boldmath Described.} The prompt specifies how to construct the run, such
as twelve copies of digit 3, and uses placeholders in the format
example. The run first occurs in the output, but the instruction makes
the digit salient.

\textbf{\boldmath Masking differential.} Two probe lines share a response, and the
choice is measured on line two. Type A repeats line one's digit, so its
context is masked; type B uses a different digit, so its context is new.
Describing the digits keeps their runs out of the prompt. Both choices
share a request and hence its key, avoiding cross-request key rotation.
Under the design's null, the second choice has the same distribution in
A and B. We sum the chi-square statistics of the \(2 \times |\Sigma|\)
type-by-word tables over contexts and permute type labels within context
(Appendix~\ref{app:A}). The \texttt{mixed} variant gives line one three
copies of one digit followed by fifteen of another: A matches line two's
trailing window and B does not, while neither repeats an entire number.

\textbf{\boldmath Spaced.} The prompt shows an arbitrary number with a space after
every digit and asks for its compact form. The compact context first
occurs in the output, without naming an individual digit in the
instruction.

\subsection{Prompt-aware reference}\label{app:A.6}

Our variant seeds the reference processor's history with hashes of the
prompt's \(h\)-token windows, retaining the most recent hashes up to the
history buffer's capacity. It also initializes the rolling context to
\texttt{prompt{[}:,\ -h-1:-1{]}} before the processor appends the
current token. Keys and the sampling rule are unchanged. Reproducing a
retained context then invokes masking. Besides defeating echo probes,
this avoids marking a user's copied text as model-generated.

\subsection{Word-list selection and safeguards}\label{app:A.7}

\emph{Reference settings.} Unless varied explicitly, SynthID-Text uses
nine keys and a four-token context (\(h = 4\)).

\emph{Word lists.} The published protocol accepts a list if its pooled
top-word share is at most 80\%, which can hide near-deterministic
choices inside cells. Fable exhibited exactly this: a pooled-balanced
list had mean per-cell top share 0.8. We instead choose the list with
highest within-cell entropy, estimated from five responses in each of
nine cells. This selects
peaches/\allowbreak{}plums/\allowbreak{}cherries/\allowbreak{}apricots,
rejected by the pooled criterion. Section~\ref{sec:4} uses a further
pilot over all ten prefixes; harvested was refused throughout.

\emph{Safeguards.} A mean per-cell top-word share above 0.9 makes a run
uninformative. Grids are scored under both \(\hat\sigma\) estimators and
both permutation nulls; only the pre-registered statistic is
confirmatory. Other scores are reported without multiplicity correction
and cannot substitute for the planned decision. Empty cells receive
their row median in both observed and permuted grids; the reference
code's shared prior can make clustered failures look like column
agreement.

\emph{Sample sizes.} Against the published paper's 100 samples per cell,
we use 5 to 13 in Section~\ref{sec:4.1}, fifty-two in
Section~\ref{sec:3.2}, and up to 65 in Section~\ref{sec:4}, with 10 000
permutations except where stated. Section~\ref{sec:3} tests sensitivity
to sample size on the reference.

\subsection{The masking-differential statistic}\label{app:A.8}

For each context \(c\) let \(n^{A}_{c,w}\) and \(n^{B}_{c,w}\) be the
number of responses of each type whose second line chose word \(w\).
With Haldane smoothing of \(1/2\) per cell, the per-context table
\(M_c = [n^{A}_{c,\cdot}; n^{B}_{c,\cdot}]\) has Pearson statistic

\[\chi^2_c = \sum_{t \in \{A,B\}} \sum_{w} \frac{ (M_c[t,w] - E_c[t,w])^2}{E_c[t,w]},\]

where \(E_c\) is the product of the row and column marginals divided by
the total, and \(T = \sum_c \chi^2_c\). The null distribution is
obtained by permuting the type labels among the responses of each
context independently and recomputing \(T\). We report the upper end of
a 99\% Clopper-Pearson interval on the proportion of permutations with
\(T\) at least the observed value. The stratified variant of
Section~\ref{sec:3.2} conditions the same procedure on the word chosen
on line one. It permutes labels within each (context, first-line word)
stratum.

\section{Tokenization and context units}\label{app:B}

\subsection{How digit length determines the context}\label{app:B.1}

The probe controls digits; the watermark uses tokens. Qwen encodes each
digit separately. Claude's token-counting endpoint instead supports
grouping digits three to a token, with separate spaces around the run
(Appendix~\ref{app:B}):

\texttt{{[}system\ prompt{]}\ {[}instruction{]}\ \textbar{}\ {[}prefix{]}\ {[}␣{]}\ {[}ddd{]}\ {[}ddd{]}\ …\ {[}partial{]}\ {[}␣{]}\ \textless{}word\textgreater{}}

The bar separates prompt and generation. Under this inferred
segmentation, a run of \(H\) digits covers \(\lceil H/3 \rceil + 2\)
tokens before the context reaches the prefix. Three consequences matter:

\begin{enumerate}
\def\labelenumi{\arabic{enumi}.}
\tightlist
\item
  A one-token watermark context sees the same trailing space in every
  cell and is invisible to the column statistic.
\item
  At \(H = 12\) the controlled context covers \(2 \leq h \leq 6\); a
  sweep over \(H \in [1,15]\) covers \(2 \leq h \leq 7\).
\item
  On Qwen a repeated-digit run stops detecting at \(H = h+1\), when
  masking applies: the top of the detected band identifies \(h\). On
  Claude, for \(h \geq 2\), the context ending in the trailing space has
  not occurred earlier: the other space follows the prefix, not digits.
  Detection can continue above the minimum length, so its bottom
  boundary gives \(\lceil H_{\min}/3 \rceil + 2\), subject to the
  segmentation and decision conditions.
\end{enumerate}

\begin{table}[tbp]
\centering
\caption{Digit runs and the token contexts they
determine.}\label{tab:8}
\vspace{\baselineskip}
\begin{tabular}{@{}
  >{\raggedright\arraybackslash}p{(\linewidth - 6\tabcolsep) * \real{0.2206}}
  >{\raggedright\arraybackslash}p{(\linewidth - 6\tabcolsep) * \real{0.2059}}
  >{\raggedright\arraybackslash}p{(\linewidth - 6\tabcolsep) * \real{0.2353}}
  >{\raggedright\arraybackslash}p{(\linewidth - 6\tabcolsep) * \real{0.3382}}@{}}
\toprule
\begin{minipage}[b]{\linewidth}\raggedright
\(H\) (digits)
\end{minipage} & \begin{minipage}[b]{\linewidth}\raggedright
Qwen tokens
\end{minipage} & \begin{minipage}[b]{\linewidth}\raggedright
Claude tokens
\end{minipage} & \begin{minipage}[b]{\linewidth}\raggedright
Claude window covers
\end{minipage} \\
\midrule
4 & 4 & 2 + 2 spaces & \(h \leq 4\) \\
6 & 6 & 2 + 2 spaces & \(h \leq 4\) \\
12 & 12 & 4 + 2 spaces & \(h \leq 6\) \\
24 & 24 & 8 + 2 spaces & \(h \leq 10\) \\
\bottomrule
\end{tabular}
\end{table}

We also test digit words of distinct lengths
(\texttt{3\ 33\ 333\ 3333\ …}), covering a word hash to depth \(H\) and
a token hash to about \(H + \sum_k \lceil k/3 \rceil\) without repeated
windows, and digit-free pronounceable non-words to test a normaliser
that drops numbers. A sweep varies \(H\), fixing the word list on the
hosted model but re-selecting it by the pooled criterion at each
reference-model length. The alternative Algorithm 1 perturbs a digit
ahead of the run to find where it stops mattering; it costs several
times more queries and is noisy by one at nine perturbation samples
(Appendix~\ref{app:C.1}).

\subsection{Token-count measurements}\label{app:B.2}

Table~\ref{tab:9} gives token counts from the \texttt{count\_tokens}
endpoint, as increments over the count for ``I ate'', measured on 6, 9
and 12 September 2026. The two models agree except on a trailing space.

\begin{table}[tbp]
\centering
\caption{Token counts on the two Claude models, as increments
over ``I ate''.}\label{tab:9}
\vspace{\baselineskip}
\begin{tabular}{@{}
  >{\raggedright\arraybackslash}p{(\linewidth - 4\tabcolsep) * \real{0.5294}}
  >{\raggedright\arraybackslash}p{(\linewidth - 4\tabcolsep) * \real{0.2353}}
  >{\raggedright\arraybackslash}p{(\linewidth - 4\tabcolsep) * \real{0.2353}}@{}}
\toprule
\begin{minipage}[b]{\linewidth}\raggedright
string appended to ``I ate''
\end{minipage} & \begin{minipage}[b]{\linewidth}\raggedright
Haiku 4.5
\end{minipage} & \begin{minipage}[b]{\linewidth}\raggedright
Fable 5.1
\end{minipage} \\
\midrule
'' 3'', '' 33'', '' 333'' & +2 & +2 \\
'' 3333'', '' 333333'' & +3 & +3 \\
'' 3333333'', '' 333333333'' & +4 & +4 \\
'' 333333333333'' & +5 & +5 \\
``3'' (no space) & +1 & +1 \\
'' '' (trailing space) & +1 & +0 \\
'' 3333 pears.'' & +7 & +7 \\
'' 3333 pears'' & +6 & +6 \\
'' 3333pears'' & +5 & +5 \\
'' pears'' & +2 & +2 \\
'' 4738291056243817905286317490263851974061 plums.'' & +19 & +19 \\
\bottomrule
\end{tabular}
\end{table}

The endpoint returns counts, which the provider documents as estimates,
not boundaries. Counting the tokens of ``I ate'' followed by the first
\(k\) digits of a number for \(k = 1, \dots, 40\), the count rises at
\(k = 1, 4, 7, \dots, 37, 40\), for the number of Table~\ref{tab:15} and
for the numbers of the inside and outside grids of Section~\ref{sec:4}
alike; this is consistent with boundaries at those digits and cannot
exclude a segmentation that changes as the string grows. Digits are
grouped three to a token from the start of the run. A space before a
digit run is its own token. The space between a digit run and the word
is its own token: '' 3333 pears'' costs one token more than ''
3333pears''. The forty-digit sentence is 19 tokens over ``I ate'': the
space, thirteen three-digit tokens, the lone last digit, the space, and
three for ``plums.''. After a word the space merges into the next word
token ('' pears'' is two tokens, the same as ``pears'' after a digit run
and its space), after a digit run it does not. Qwen2.5 emits one token
per digit, keeps a separate space token before a digit run, and merges
the space into a following word.

\section{Reference validation and power}\label{app:C}

We compare the same Qwen2.5-3B weights unwatermarked, with stock
SynthID-Text, and with the prompt-aware variant. Unless stated
otherwise, \(H = h = 4\), each cell has ten samples, and the list is
peaches/\allowbreak{}plums/\allowbreak{}cherries/\allowbreak{}apricots.
The described probe instead uses
mangoes/\allowbreak{}pineapples/\allowbreak{}papayas/\allowbreak{}kiwis
because its wording collapses the first list onto peaches.

\subsection{The Red-Green test on SynthID-Text}\label{app:C.1}

The published probe detects the reference and stays clean on the control
(Table~\ref{tab:10}). The code's estimator gives the maximum \(S = 10\);
the paper's estimator treats some column shift as noise, raising the
threshold but retaining detection.

\begin{table}[tbp]
\centering
\caption{The published probe on the local reference. Ten
samples per cell, 10 000 permutations.}\label{tab:10}
\vspace{\baselineskip}
\begin{tabular}{@{}
  >{\raggedright\arraybackslash}p{(\linewidth - 8\tabcolsep) * \real{0.2135}}
  >{\raggedright\arraybackslash}p{(\linewidth - 8\tabcolsep) * \real{0.2809}}
  >{\raggedright\arraybackslash}p{(\linewidth - 8\tabcolsep) * \real{0.0787}}
  >{\raggedright\arraybackslash}p{(\linewidth - 8\tabcolsep) * \real{0.2022}}
  >{\raggedright\arraybackslash}p{(\linewidth - 8\tabcolsep) * \real{0.2247}}@{}}
\toprule
\begin{minipage}[b]{\linewidth}\raggedright
model
\end{minipage} & \begin{minipage}[b]{\linewidth}\raggedright
\(\hat\sigma\) estimator
\end{minipage} & \begin{minipage}[b]{\linewidth}\raggedright
\(S\)
\end{minipage} & \begin{minipage}[b]{\linewidth}\raggedright
\(p\) (99\% upper)
\end{minipage} & \begin{minipage}[b]{\linewidth}\raggedright
verdict
\end{minipage} \\
\midrule
SynthID, \(H = 4\) & code & 10 & 0.0005 & detected \\
SynthID, \(H = 4\) & paper & 5 & 0.0047 & detected \\
control, \(H = 4\) & code & 1 & 1.00 & clean \\
SynthID, \(H = 5\) & code & 0 & 1.00 & hidden by masking \\
\bottomrule
\end{tabular}
\end{table}

\begin{figure}
\centering
\includegraphics[width=1\linewidth,height=\textheight,keepaspectratio,alt={Permutation p-value against the length of the digit run. On the reference model detection starts at H = 3, where a constant space token completes the four-token window, and stops at H = 5, where the window repeats and the watermark is masked. Haiku 4.5 is flat.}]{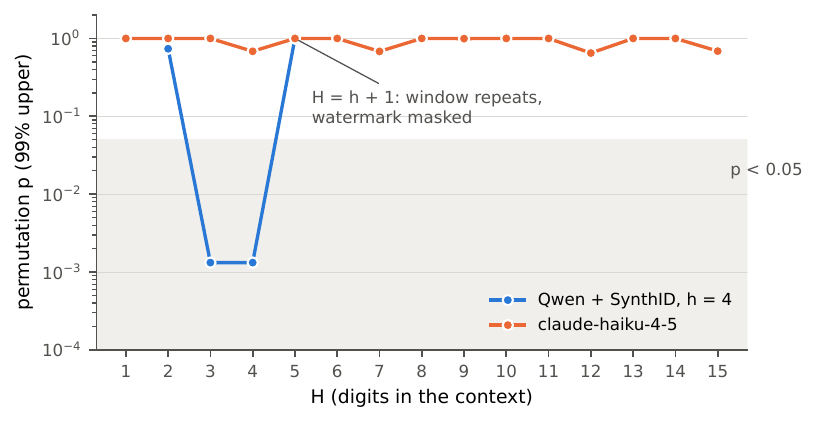}
\caption{Permutation \(p\)-value against the length of the digit run. On
the reference model detection starts at \(H = 3\), where a constant
space token completes the four-token window, and stops at \(H = 5\),
where the window repeats and the watermark is masked. Haiku 4.5 is
flat.}\label{fig:3}
\end{figure}

The last row is the masking rule at work: at five digits the window
before the choice repeats the one before it, and the test sees nothing,
the constraint \citet[App. F.4]{gloaguen2025} report. A sweep over
\(H \in [2, 5]\) therefore detects at \(H = 3\) and \(4\) only, and
returns \(\hat h = 4\). The hit at \(H = 3\) comes from the constant
space token, which fills the fourth slot. The paper's Algorithm 1
perturbs a digit ahead of the run and looks for the length at which the
perturbation stops mattering. It also returns \(\hat h = 4\) under the
rule stated in the paper, which compares consecutive lengths. Under the
rule in the released code, which compares every length against the
shortest, it returns \(\hat h = 5\). Figure~\ref{fig:3} plots the sweep
next to the Haiku sweep of Section~\ref{sec:4.1}.

\begin{figure}
\centering
\includegraphics[width=1\linewidth,height=\textheight,keepaspectratio,alt={Detection and false-positive rates from resampling the reference matrices, over 30 trials: against samples per cell on the full grid (left) and against grid size at ten samples per cell (right).}]{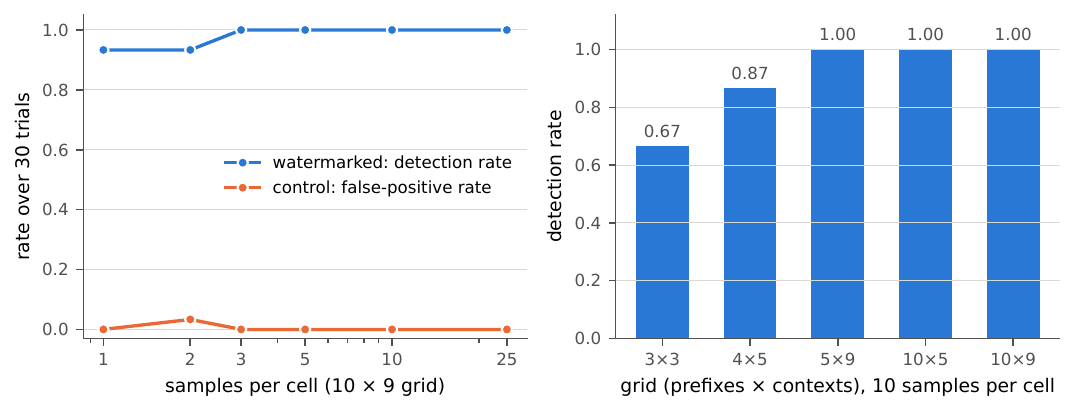}
\caption{Detection and false-positive rates from resampling the
reference matrices, over 30 trials: against samples per cell on the full
grid (left) and against grid size at ten samples per cell
(right).}\label{fig:4}
\end{figure}

Sample size is not the binding constraint at this strength. We resampled
the collected matrices at smaller \(N\). The test detects in 30 of 30
trials at three samples per cell, and in 28 of 30 at one. Reducing the
grid instead, to \(3 \times 3\) cells, drops detection to about two
thirds of trials even with exact probabilities. The reason is that the
permutation null has too few arrangements to form a tail.
Appendix~\ref{app:C.3} extends this to weaker watermarks and sharper
choices. The hosted runs of Section~\ref{sec:4.1} use between five and
thirteen samples per cell on the full grid (Figure~\ref{fig:4}).

\subsection{Three probes against three ground truths}\label{app:C.2}

Table~\ref{tab:1} runs each probe of Section~\ref{sec:2.2} against each
of the three models. Every entry is a separate experiment on the same
weights.

\textbf{\boldmath The published probe goes blind.}

\begin{figure}
\centering
\includegraphics[width=1\linewidth,height=\textheight,keepaspectratio,alt={Row-centred logit of the preferred word for every cell, H = 4 locally and H = 12 on Fable, red below and green above as the test colours them. Outlined cells exceed the flag threshold 1.96\textbackslash hat\textbackslash sigma. Only the reference model shows columns; the prompt-aware model carries the same watermark and shows none.}]{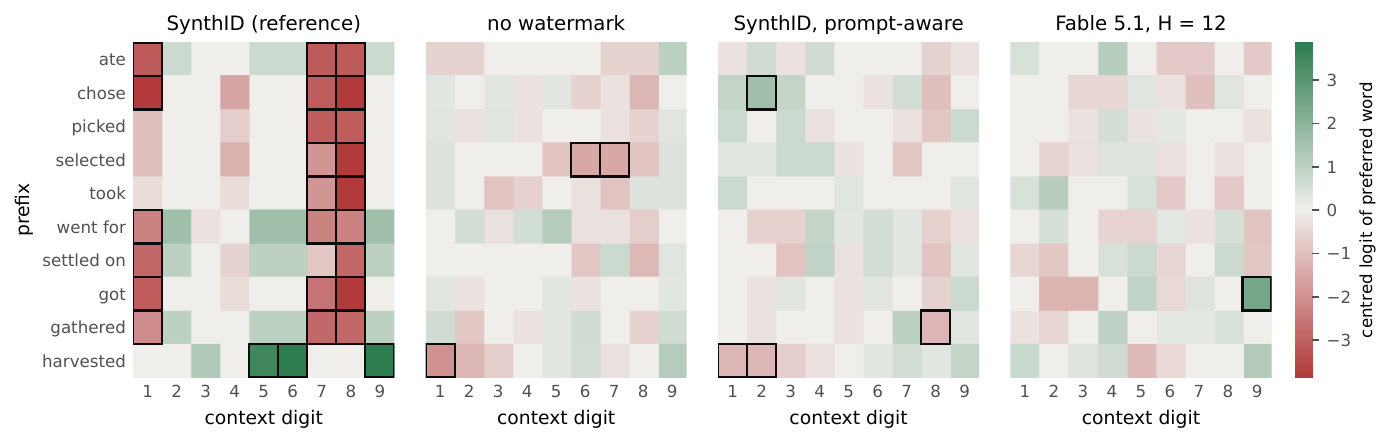}
\caption{Row-centred logit of the preferred word for every cell,
\(H = 4\) locally and \(H = 12\) on Fable, red below and green above as
the test colours them. Outlined cells exceed the flag threshold
\(1.96\hat\sigma\). Only the reference model shows columns; the
prompt-aware model carries the same watermark and shows
none.}\label{fig:5}
\end{figure}

The prompt-aware initialization makes the echo probe indistinguishable
from the clean control, although the model retains its watermark
(Table~\ref{tab:1}). The measured continuation is unmarked by
construction, so additional samples cannot recover a signal there.
Figure~\ref{fig:5} compares the matrices with the deployment's null
grid.

\textbf{\boldmath The described probe restores the signal and adds a confound.}

Both watermarked variants are detected, but so is the clean control:
naming a digit in the instruction creates a digit-to-fruit preference.
On the clean model, variation in mean choice across digits is twelve
times larger under described than echo wording (0.094 versus 0.008;
permutation \(p < 10^{-3}\) versus \(p = 0.08\)). Three digits change
the preferred word. This native column structure mimics a watermark. The
paper's estimator stays clean (\(S = 2\), \(p = 0.60\)) but also loses
detection on the watermarked variants.

\textbf{\boldmath The masking differential is correct in all three columns.} It
detects both watermark variants and stays clean on the control.
Context-specific preferences match the predicted masking effect: after
7777, type A chooses cherries in all ten samples and type B plums in all
ten. Sharing a request also protects the comparison against
cross-request key rotation.

\subsection{Power against watermark strength and choice
entropy}\label{app:C.3}

\begin{figure}
\centering
\includegraphics[width=1\linewidth,height=\textheight,keepaspectratio,alt={Left: detection against samples per cell for each number of keys. Right: detection against per-cell entropy at nine keys, with the range observed on Fable shaded and the control's false-positive rate marked.}]{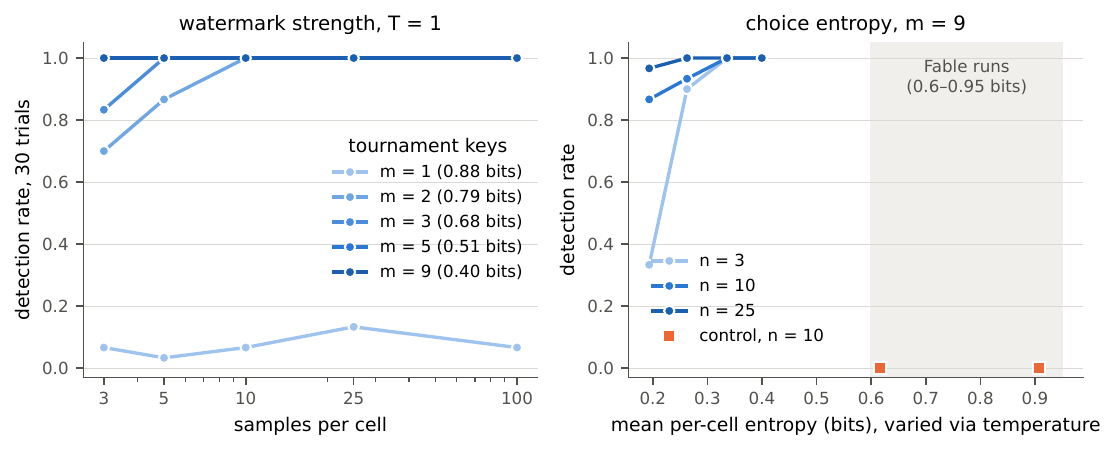}
\caption{Left: detection against samples per cell for each number of
keys. Right: detection against per-cell entropy at nine keys, with the
range observed on Fable shaded and the control's false-positive rate
marked.}\label{fig:6}
\end{figure}

Two quantities the field cannot control set the test's power: the
strength of the watermark and the uncertainty of the choice inside a
cell. On the reference model we vary the first through the number of
tournament keys \(m\), from one to the default nine, and the second
through the sampling temperature \(\tau\) at \(m = 9\), measured as the
mean per-cell entropy after watermarking, the quantity we can also
measure on Fable. The exact choice distribution of every cell is read
once and black-box sampling at \(n\) draws per cell is simulated by
resampling, thirty trials per point (Figure~\ref{fig:6},
Table~\ref{tab:11}).

\begin{table}[tbp]
\centering
\caption{Detection rate of the echo probe on the reference
model against the number of tournament keys and the sampling
temperature; false-positive rate on the control in the last two rows.
Thirty resampling trials per entry, 1 000 permutations each. For one key
the rate is 0.03 at 300 and 0.13 at 1 000 samples per cell. With thirty
trials a rate of 0 has a 95\% upper bound of about 0.12.}\label{tab:11}
\vspace{\baselineskip}
\small\setlength{\tabcolsep}{3pt}
\begin{tabular}{@{}
  >{\raggedright\arraybackslash}p{(\linewidth - 16\tabcolsep) * \real{0.1068}}
  >{\raggedright\arraybackslash}p{(\linewidth - 16\tabcolsep) * \real{0.0874}}
  >{\raggedright\arraybackslash}p{(\linewidth - 16\tabcolsep) * \real{0.1650}}
  >{\raggedright\arraybackslash}p{(\linewidth - 16\tabcolsep) * \real{0.1165}}
  >{\raggedright\arraybackslash}p{(\linewidth - 16\tabcolsep) * \real{0.0971}}
  >{\raggedright\arraybackslash}p{(\linewidth - 16\tabcolsep) * \real{0.0971}}
  >{\raggedright\arraybackslash}p{(\linewidth - 16\tabcolsep) * \real{0.1068}}
  >{\raggedright\arraybackslash}p{(\linewidth - 16\tabcolsep) * \real{0.1068}}
  >{\raggedright\arraybackslash}p{(\linewidth - 16\tabcolsep) * \real{0.1165}}@{}}
\toprule
\begin{minipage}[b]{\linewidth}\raggedright
keys \(m\)
\end{minipage} & \begin{minipage}[b]{\linewidth}\raggedright
\(\tau\)
\end{minipage} & \begin{minipage}[b]{\linewidth}\raggedright
entropy (bits)
\end{minipage} & \begin{minipage}[b]{\linewidth}\raggedright
top share
\end{minipage} & \begin{minipage}[b]{\linewidth}\raggedright
\(n = 3\)
\end{minipage} & \begin{minipage}[b]{\linewidth}\raggedright
\(n = 5\)
\end{minipage} & \begin{minipage}[b]{\linewidth}\raggedright
\(n = 10\)
\end{minipage} & \begin{minipage}[b]{\linewidth}\raggedright
\(n = 25\)
\end{minipage} & \begin{minipage}[b]{\linewidth}\raggedright
\(n = 100\)
\end{minipage} \\
\midrule
1 & 1.0 & 0.88 & 0.72 & 0.07 & 0.03 & 0.07 & 0.13 & 0.07 \\
2 & 1.0 & 0.79 & 0.76 & 0.70 & 0.87 & 1.00 & 1.00 & 1.00 \\
3 & 1.0 & 0.68 & 0.80 & 0.83 & 1.00 & 1.00 & 1.00 & 1.00 \\
5 & 1.0 & 0.51 & 0.87 & 1.00 & 1.00 & 1.00 & 1.00 & 1.00 \\
9 & 1.0 & 0.40 & 0.89 & 1.00 & 1.00 & 1.00 & 1.00 & 1.00 \\
9 & 0.8 & 0.34 & 0.91 & 1.00 & 1.00 & 1.00 & 1.00 & 1.00 \\
9 & 0.6 & 0.26 & 0.93 & 0.90 & 0.97 & 0.93 & 1.00 & 1.00 \\
9 & 0.45 & 0.19 & 0.95 & 0.33 & 0.90 & 0.87 & 0.97 & 1.00 \\
none & 1.0 & 0.91 & 0.72 & 0.00 & 0.00 & 0.00 & 0.00 & 0.00 \\
none & 0.6 & 0.62 & 0.81 & 0.00 & 0.00 & 0.00 & 0.00 & 0.00 \\
\bottomrule
\end{tabular}
\end{table}

The test has a floor in watermark strength that more queries do not
lower. With one key it detects in 7\% of trials at 100 samples per cell
and 13\% at 1 000; two keys suffice for detection in every trial at ten
samples. The threshold is structural: \(\hat\sigma\) measures variation
across prefixes and does not shrink with \(n\), leaving a bias below
about \(1.96\hat\sigma\) unflagged.

Per-cell entropy alone does not explain the field nulls: at 0.26 bits
and ten samples the reference is detected in 93\% of trials, below
Fable's 0.6 to 0.95 bits. The deployment's unknown per-step bias remains
a separate limitation.

\section{Deployment configurations and differential
diagnostics}\label{app:D}

\subsection{Haiku sweep}\label{app:D.1}

The Haiku echo sweep over \(H \in [1,15]\), at ten samples per cell with
every response parsed, finds no Red-Green structure over contexts of two
to seven tokens (Figure~\ref{fig:3}; Appendix~\ref{app:D}). The code's
estimator gives \(S \leq 2\) and \(p \geq 0.64\) throughout; the paper's
reaches \(p = 0.06\)-\(0.12\) at three lengths without rejecting.

Table~\ref{tab:12} gives the sweep of Section~\ref{sec:4.1} at every
length.

\begin{table}[tbp]
\centering
\caption{The echo probe on \protect\texttt{claude-haiku-4-5}
at digit lengths 1 to 15, ten samples per cell, no refusals; \(p\) is
the 99\% upper bound under each estimator.}\label{tab:12}
\vspace{\baselineskip}
\begin{tabular}{@{}
  >{\raggedright\arraybackslash}p{(\linewidth - 8\tabcolsep) * \real{0.0805}}
  >{\raggedright\arraybackslash}p{(\linewidth - 8\tabcolsep) * \real{0.3103}}
  >{\raggedright\arraybackslash}p{(\linewidth - 8\tabcolsep) * \real{0.1609}}
  >{\raggedright\arraybackslash}p{(\linewidth - 8\tabcolsep) * \real{0.2989}}
  >{\raggedright\arraybackslash}p{(\linewidth - 8\tabcolsep) * \real{0.1494}}@{}}
\toprule
\begin{minipage}[b]{\linewidth}\raggedright
\(H\)
\end{minipage} & \begin{minipage}[b]{\linewidth}\raggedright
\(S\) (paper \(\hat\sigma\))
\end{minipage} & \begin{minipage}[b]{\linewidth}\raggedright
\(p\) (paper)
\end{minipage} & \begin{minipage}[b]{\linewidth}\raggedright
\(S\) (code \(\hat\sigma\))
\end{minipage} & \begin{minipage}[b]{\linewidth}\raggedright
\(p\) (code)
\end{minipage} \\
\midrule
1 & 2 & 0.67 & 1 & 1.00 \\
2 & 1 & 1.00 & 1 & 1.00 \\
3 & 2 & 0.60 & 0 & 1.00 \\
4 & 2 & 0.71 & 2 & 0.67 \\
5 & 1 & 1.00 & 1 & 1.00 \\
6 & 1 & 1.00 & 1 & 1.00 \\
7 & 2 & 0.71 & 2 & 0.67 \\
8 & 1 & 1.00 & 1 & 1.00 \\
9 & 3 & 0.06 & 1 & 1.00 \\
10 & 3 & 0.10 & 0 & 1.00 \\
11 & 2 & 0.67 & 1 & 1.00 \\
12 & 3 & 0.12 & 2 & 0.64 \\
13 & 1 & 1.00 & 1 & 1.00 \\
14 & 1 & 1.00 & 1 & 1.00 \\
15 & 2 & 0.73 & 2 & 0.68 \\
\bottomrule
\end{tabular}
\end{table}

\subsection{Initial Fable configurations}\label{app:D.2}

Table~\ref{tab:13} lists the ten initial Fable configurations in
collection order; Table~\ref{tab:14} maps their nulls to alternative
explanations. Long-context pilots and planned runs follow in
Section~\ref{sec:4} and Appendix~\ref{app:H}. Here \(p\) and \(p'\) are
the 99\% permutation upper bounds under the code's and paper's
\(\hat\sigma\) estimators.

\begin{table}[tbp]
\centering
\caption{All configurations on
\protect\texttt{claude-fable-5-1}. \(n\) is samples per cell after
refusals; ``uninf.'' is the safeguard of Section~\ref{sec:2.3}. Scored
by the collection script, with one pseudo-count per word and an empty
cell at its row median. ``peaches'' is
peaches/\allowbreak{}plums/\allowbreak{}cherries/\allowbreak{}apricots,
``apples'' is
apples/\allowbreak{}bananas/\allowbreak{}oranges/\allowbreak{}pears.}\label{tab:13}
\vspace{\baselineskip}
\footnotesize\setlength{\tabcolsep}{2.5pt}
\begin{tabular}{@{}
  >{\raggedright\arraybackslash}p{(\linewidth - 20\tabcolsep) * \real{0.0560}}
  >{\raggedright\arraybackslash}p{(\linewidth - 20\tabcolsep) * \real{0.1920}}
  >{\raggedright\arraybackslash}p{(\linewidth - 20\tabcolsep) * \real{0.0960}}
  >{\raggedright\arraybackslash}p{(\linewidth - 20\tabcolsep) * \real{0.1600}}
  >{\raggedright\arraybackslash}p{(\linewidth - 20\tabcolsep) * \real{0.0960}}
  >{\raggedright\arraybackslash}p{(\linewidth - 20\tabcolsep) * \real{0.0800}}
  >{\raggedright\arraybackslash}p{(\linewidth - 20\tabcolsep) * \real{0.0560}}
  >{\raggedright\arraybackslash}p{(\linewidth - 20\tabcolsep) * \real{0.0560}}
  >{\raggedright\arraybackslash}p{(\linewidth - 20\tabcolsep) * \real{0.0560}}
  >{\raggedright\arraybackslash}p{(\linewidth - 20\tabcolsep) * \real{0.0560}}
  >{\raggedright\arraybackslash}p{(\linewidth - 20\tabcolsep) * \real{0.0960}}@{}}
\toprule
\begin{minipage}[b]{\linewidth}\raggedright
\#
\end{minipage} & \begin{minipage}[b]{\linewidth}\raggedright
probe
\end{minipage} & \begin{minipage}[b]{\linewidth}\raggedright
transport
\end{minipage} & \begin{minipage}[b]{\linewidth}\raggedright
context
\end{minipage} & \begin{minipage}[b]{\linewidth}\raggedright
list
\end{minipage} & \begin{minipage}[b]{\linewidth}\raggedright
\(H\)
\end{minipage} & \begin{minipage}[b]{\linewidth}\raggedright
\(n\)
\end{minipage} & \begin{minipage}[b]{\linewidth}\raggedright
\(S\)
\end{minipage} & \begin{minipage}[b]{\linewidth}\raggedright
\(p\)
\end{minipage} & \begin{minipage}[b]{\linewidth}\raggedright
\(p'\)
\end{minipage} & \begin{minipage}[b]{\linewidth}\raggedright
result
\end{minipage} \\
\midrule
1 & echo & batch & digits, spelled & peaches & 12 & 6 & 1 & 1.00 & 0.90
& null \\
2a & echo & batch & digits, spelled & apples & 6 & 13 & 1 & 0.99 & 0.99
& null \\
2b & echo & batch & digits, spelled & apples & 12 & 13 & 1 & 1.00 & 0.51
& null \\
2c & echo & batch & digits, spelled & apples & 24 & 13 & 0 & 1.00 & 0.71
& null \\
A & echo & batch & six digit words & apples & 6 words & 7 & 1 & 1.00 &
0.13 & null \\
B & echo, 90-word paragraph first & batch & six digit words & apples & 6
words & 7 & 4 & 0.43 & 0.43 & uninf. \\
P1 & described & batch & digits, constructed & apples & 12 & ≤ 8 & 0 &
1.00 & 0.07 & null \\
S & echo & sync & digits, spelled & peaches & 12 & 5 & 1 & 1.00 & 0.89 &
null \\
B2 & echo, 150-word paragraph first, \texttt{max\_tokens} 1024 & batch &
digits, spelled & peaches & 12 & 6 & 1 & 1.00 & 1.00 & null \\
W & echo & batch & four nonsense words & peaches & 4 words & ≤ 8 & 2 &
0.54 & 0.57 & null \\
\bottomrule
\end{tabular}
\end{table}

In every run the context was reproduced intact in 100\% of parsed
responses; per-cell entropy, measured in runs 2, P1, S, B2 and W, was
between 0.6 and 0.95 bits. Run B is uninformative by the safeguard of
Section~\ref{sec:2.3} (``pears'' in 612 of 628 responses). The refusals
in P1 and W, 11\% each, left seven and five of the ninety cells empty;
an empty cell takes its row median and cannot add column structure. The
test never rejects; the largest statistic is \(S = 2\), where the
reference model at the same sample size gave \(S = 9\)--\(10\).
Table~\ref{tab:14} states what each run excludes; three explanations
remain that no probe here can address, a key that changes between
requests, a hash over more than the local window, and marking that is
inactive on this route, and Section~\ref{sec:3.2} and
Section~\ref{sec:4} return to them.

\begin{table}[tbp]
\centering
\caption{Explanations for the Fable null and the runs that
bear on them. Rows marked ``rejected'' on the strength of echo probes
(A, 2, W, S, B2) hold for a scheme with the reference's prompt-blind
history. The deployment contrast is consistent with prompt-aware history
(Section~\ref{sec:4}), so these exclusions cannot be assumed to apply to
it. Runs P24, P40 and W40b are described in Section~\ref{sec:4} and
Appendix~\ref{app:H}, and the last five rows are discussed
there.}\label{tab:14}
\vspace{\baselineskip}
\begin{tabular}{@{}
  >{\raggedright\arraybackslash}p{(\linewidth - 4\tabcolsep) * \real{0.3000}}
  >{\raggedright\arraybackslash}p{(\linewidth - 4\tabcolsep) * \real{0.5200}}
  >{\raggedright\arraybackslash}p{(\linewidth - 4\tabcolsep) * \real{0.1800}}@{}}
\toprule
\begin{minipage}[b]{\linewidth}\raggedright
hypothesis for the null
\end{minipage} & \begin{minipage}[b]{\linewidth}\raggedright
status
\end{minipage} & \begin{minipage}[b]{\linewidth}\raggedright
run
\end{minipage} \\
\midrule
prompt-aware masking alone, window up to 16 tokens & rejected as sole
cause; consistent with the echoed grid E40 (Section~\ref{sec:4}) & P1,
P24, P40, E40 \\
word-level hash, up to six words & rejected for a prompt-blind scheme &
A \\
token context up to fifteen tokens & rejected for a prompt-blind scheme
& A, 2 \\
hash that ignores digits & rejected for a prompt-blind scheme, word
contexts up to four words & W \\
reasoning tokens ahead of the answer & rejected: none emitted & all \\
batch transport not yet marked & rejected for a prompt-blind scheme &
S \\
marking gated on output length or \texttt{max\_tokens} & rejected for a
prompt-blind scheme, up to \textasciitilde200 tokens & B2 \\
insufficient entropy or samples & addressed: per-cell selection;
decisive locally at the same \(n\) and lower entropy
(Appendix~\ref{app:C.3}) & --- \\
per-step bias as weak as a one-key tournament & open: low detection in
the tested configuration up to 1 000 samples per cell
(Appendix~\ref{app:C.3}) & --- \\
key varies across requests & rejected for a fresh key per request, if
the shift of Section~\ref{sec:4} is the watermark; a small key pool
keeps an average bias \citep[App. F.2]{gloaguen2025} & W40b \\
hash over more than the local window & bounded if the shift is a hash
and the segmentation of Appendix~\ref{app:B} holds: at most fourteen
tokens (Section~\ref{sec:4}, Table~\ref{tab:6}) & W40b \\
long context, assembled by the model, within-prefix
null & primary rejects, plan inconclusive (refusal
criterion); controls clean while unmarked, Sonnet 5 acquires the pattern
once marked (Sections 4.2, 4.4) & W40b \\
not active on this route & rejected on the same condition: a
context-keyed shift is present on this route (Section~\ref{sec:4}) &
W40b \\
native preference of Fable for the last digit &
inconclusive under the primary score; a new base
gives a new pattern under the secondary (Section~\ref{sec:4.3}); an
unwatermarked reference shows a shift of this size
(Section~\ref{sec:3.3}); the pattern appeared on Sonnet 5 on marking
(Section~\ref{sec:4.4}) & W40c, inside, outside \\
\bottomrule
\end{tabular}
\end{table}

\subsection{Differential diagnostics on Claude}\label{app:D.3}

The masking differential of Section~\ref{sec:2.2} was run on Fable and,
as a control, on Haiku, with fifty-two samples per cell. We ran the
original variant at \(H = 12\) and the \texttt{mixed} variant at
\(H = 18\). Table~\ref{tab:2} shows that it rejects in all four cases.

\begin{figure}
\centering
\includegraphics[width=1\linewidth,height=\textheight,keepaspectratio,alt={The fruit chosen on line one, before any context can repeat, under the two request types. Both models shift it according to whether the prompt asks for the same number twice; the difference is what the masking-differential statistic then picks up on line two.}]{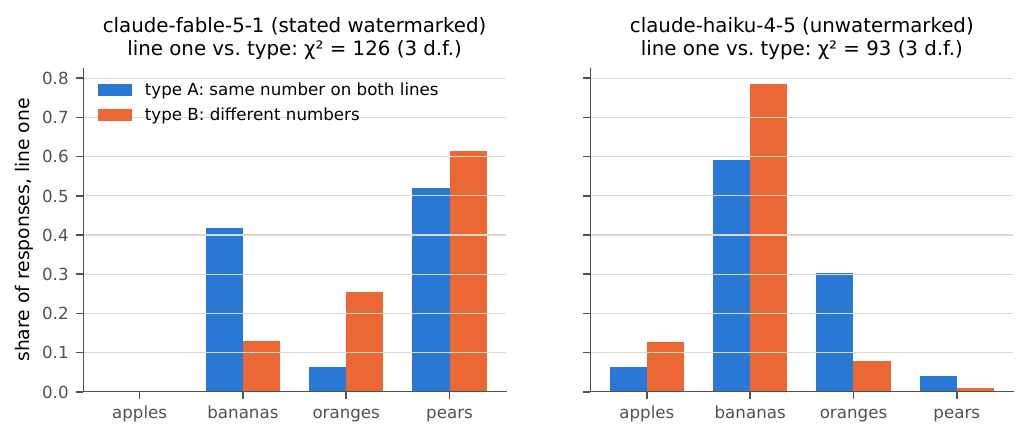}
\caption{The fruit chosen on line one, before any context can repeat,
under the two request types. Both models shift it according to whether
the prompt asks for the same number twice; the difference is what the
masking-differential statistic then picks up on line two.}\label{fig:7}
\end{figure}

The test rejects the null on the unwatermarked model as strongly as on
the watermarked one. The responses reveal a confound: neither model ever
repeats the first fruit on line two, and the first fruit already depends
on the request type. Fable chooses bananas in 42\% of type A and 13\% of
type B responses, despite matched pooled first-line digits. Asking for
the same number twice or for different numbers changes the model's first
choice before any context can repeat; the no-repetition habit carries
that difference into line two (Figure~\ref{fig:7}).

The cue also acts on the second line directly, not only through the
first: conditioning on the first fruit leaves residual type effects on
both models (summed within-fruit chi-square, type labels permuted within
strata: Fable \(T = 42.5\), Haiku \(T = 71.7\), both
\(p_{99} = 0.003\)). The \texttt{mixed} variant reduces but does not
remove the first-line dependence, and the control still rejects. Thus
the local clean control does not transfer to Claude. A potential repair,
not tested here, is to give both conditions the same prompt, let the
model choose line one's trailing digit, and classify responses
afterwards.

\section{Main-grid sensitivity, follow-ups and window
search}\label{app:E}

The final design keeps a long compact context out of the prompt while
making its variants minimally different in meaning: a forty-digit number
shown with spaces between digits, reproduced compactly by the model,
with nine contexts sharing the first thirty-nine digits. Its inferred
window is sixteen tokens including spaces. To accommodate native prefix
effects, the \emph{within-prefix score} retains Appendix~\ref{app:A.2}'s
column statistic, caps cell rates away from 0 and 1 before the logit,
and permutes contexts within each prefix. Appendix~\ref{app:H} checks
the design and null on reference and Claude grids.

Three Fable and three Haiku pilots at ten to thirteen samples per cell
were null under this score. They also showed that a within-prefix
chi-square rejects on unwatermarked Claude and that the apples list
fixes half of Fable's prefixes on one word. We then fixed a plan before
collection, without external timestamping (deviations in
Appendix~\ref{app:I}): choose a list by within-prefix entropy; collect
fifty samples per cell, with a same-day Haiku control; use the
within-prefix score, code's \(\hat\sigma\), and 10 000 permutations as
primary. A detection requires Fable \(p_{99}<0.01\), Haiku
\(p_{99}>0.05\), nondegenerate grids and refusals not concentrated by
context. The pilot selected
plums/\allowbreak{}limes/\allowbreak{}dates/\allowbreak{}kiwis at 0.99
bits per prefix, versus under 0.4 for the apples and berry lists.

\begin{table}[tbp]
\centering
\caption{The pre-registered run: spaced 40-digit probe,
contexts differing in the last digit, list
plums/\allowbreak{}limes/\allowbreak{}dates/\allowbreak{}kiwis, fifty
samples per cell planned on Fable (1 to 65 received after refusals and
retries) and sixty-five on Haiku; the Sonnet 5 control, added under a
later addendum to the plan, uses the eight well-sampled prefixes at
twenty-five samples per cell. ``Primary'' is the within-prefix score
with the code's \(\hat\sigma\); the next column is the same score with
the paper's estimator; the last is the published statistic with the
code's and then the paper's estimator, the order kept in
Table~\ref{tab:6}. 10 000 permutations. The discrete
scores in Tables 4, 5, 6 and 15 are computed by
\protect\texttt{rowwise\_stats.py}: the published rows add one
pseudo-count per word, as the collection script does, and a cell without
responses takes its row median in the observed and the permuted grids
alike.}\label{tab:15}
\vspace{\baselineskip}
\begin{tabular}{@{}
  >{\raggedright\arraybackslash}p{(\linewidth - 12\tabcolsep) * \real{0.0973}}
  >{\raggedright\arraybackslash}p{(\linewidth - 12\tabcolsep) * \real{0.1062}}
  >{\raggedright\arraybackslash}p{(\linewidth - 12\tabcolsep) * \real{0.0885}}
  >{\raggedright\arraybackslash}p{(\linewidth - 12\tabcolsep) * \real{0.1062}}
  >{\raggedright\arraybackslash}p{(\linewidth - 12\tabcolsep) * \real{0.2124}}
  >{\raggedright\arraybackslash}p{(\linewidth - 12\tabcolsep) * \real{0.2035}}
  >{\raggedright\arraybackslash}p{(\linewidth - 12\tabcolsep) * \real{0.1858}}@{}}
\toprule
\begin{minipage}[b]{\linewidth}\raggedright
model
\end{minipage} & \begin{minipage}[b]{\linewidth}\raggedright
\(n\) total
\end{minipage} & \begin{minipage}[b]{\linewidth}\raggedright
refused
\end{minipage} & \begin{minipage}[b]{\linewidth}\raggedright
top share
\end{minipage} & \begin{minipage}[b]{\linewidth}\raggedright
primary \(S\), \(p_{99}\)
\end{minipage} & \begin{minipage}[b]{\linewidth}\raggedright
paper's \(\hat\sigma\)
\end{minipage} & \begin{minipage}[b]{\linewidth}\raggedright
published \(p_{99}\)
\end{minipage} \\
\midrule
Fable & 4 965 & 44\% & 0.64 & \textbf{\boldmath 4, 0.0005} & 4, 0.055 & 0.016 /
0.008 \\
Haiku & 5 850 & 0 & 0.46 & 1, 1.00 & 3, 0.52 & 1.00 / 0.11 \\
Sonnet 5 & 2 304 & 0 & 0.75 & 1, 1.00 & 1, 1.00 & 1.00 / 1.00 \\
\bottomrule
\end{tabular}
\end{table}

Unwatermarked Claude has digit preferences within prefixes
(Appendix~\ref{app:H}). The within-prefix null retains those row effects
and tests whether their directions align across prefixes. Fable rejects
at the primary's floor, while the same score with the paper's estimator
gives \(p_{99}=0.055\). Permutation increases the code's \(\hat\sigma\),
reducing flags; it leaves the paper's estimator unchanged. The primary
\(p\)-value therefore belongs to the whole procedure, not the flag count
alone. The published scores also reject (Table~\ref{tab:15}); the capped
grid-wide scores give 0.019 and 0.013. Haiku does not reject, and
neither grid is degenerate.

The pre-registered decision is nevertheless \textbf{\boldmath inconclusive}:
first-round refusals are concentrated by context (78 to 177 by final
digit, chi-square \(p<10^{-10}\)), violating the plan. Exploratory
checks reduce but do not exclude the possibility that filtering
manufactures the alignment. Refusals return no text and, from the
echoed-grid comparison, appear prompt-dependent (Appendix~\ref{app:G});
cell refusal rate is uncorrelated with row-centred share (\(r=-0.03\),
ninety cells). The seventy-three cells refused under 30\% retain the
pattern (range 0.55, \(p_{99}=0.0007\); \texttt{analysis/refusals.py}).
Its watermark origin remains unresolved.

The effect is substantial: pooled kiwis shares after final digits 1 and
2 are 68\% and 72\%, versus 20\%, 19\% and 24\% after 3, 5 and 6.
Alternating responses into halves gives column-mean correlation 0.99.
Excluding harvested and gathered, depleted by refusals, leaves the
primary unchanged; the paper's estimator gives \(S=3\), \(p_{99}=0.22\).
To equalise cell sizes, \texttt{analysis/equal\_n.py} draws twenty-two
first-round verbatim responses per cell without replacement over these
eight prefixes, one hundred times (seeds 0 to 99; 1 500 permutations per
draw; preferred word re-selected). The primary rejects at 0.05 in 67
draws and at 0.01 in 51, while the continuous range stays at 0.50 to
0.66 throughout.

That range is the maximum minus minimum column mean of the row-centred
preferred-word share: 0.57 on Fable (percentile bootstrap 0.53 to 0.61,
biased upward for a range), versus Haiku's 0.07 (0.06 to 0.14). Its
within-prefix \(p_{99}\) is 0.0005 and 0.78, respectively, showing that
the effect does not depend on the discreteness of \(S\).

Haiku's column means correlate with Fable's at \(-0.17\). It copies the
number incorrectly in 55\% of responses; restricting to verbatim copies
remains null (\(p_{99}=1.00\)). Because the smaller, older Haiku may
differ behaviourally, we added Sonnet under a plan addendum using the
same probe, number and list. It has no refusals and thirty-two responses
per cell, stays null under every column score (Table~\ref{tab:15}; range
0.12, \(p_{99}=0.20\)), and has nonsignificant column correlation
\(+0.53\) with Fable over nine columns. Its within-prefix chi-square
rejects, as Haiku's does. With 20\% nonverbatim responses removed,
Sonnet's primary gives \(S=2\), \(p_{99}=0.08\).

The score has a reference positive control; at
submission the final spaced probe had none on a prompt-aware reference,
which Section~\ref{sec:3.3} now supplies (Appendix~\ref{app:H}).
Together with two same-family controls, the results establish a Fable
context-dependent shift; a provider detector could establish whether the
responses carry a mark, but attributing the shift to it requires further
evidence (Sections 4.4 and 4.5).

\textbf{\boldmath Follow-up grids.}

The addendum grids in Table~\ref{tab:5} test the shift's dependence on
context (designs and decision rules in Appendix~\ref{app:H}). E40 echoes
the compact number from the prompt and is null, consistent with
prompt-aware masking; number salience and filtering also differ, so this
comparison does not identify the mechanism. W40c changes the base number
and is inconclusive under the plan, although its exploratory range
rejects and its column pattern differs from W40b's (\(r=-0.34\)).
Holding the last digit fixed and varying a triple inside or outside the
proposed window gives two more primary nulls. Conspicuous triples
confound their continuous scores, as the same-family controls in
Appendix~\ref{app:H} show.

\textbf{\boldmath How long is the window.}

If the effect is a context hash, detection at digit length \(D\)
requires \(h\leq\lceil D/3\rceil+2\) under the inferred segmentation. We
searched lengths \(D=3k+1\) by planned bisection, keeping the final
digit a lone token: eight prefixes, twenty-five samples per cell, the
same list, and the primary score as the decision statistic.
Table~\ref{tab:6} reports every score.

\begin{figure}
\centering
\includegraphics[width=1\linewidth,height=\textheight,keepaspectratio,alt={The context-length search: primary p\_\{99\} against the window before the choice. Hollow points are the grids that the safety classifier depleted unevenly (D = 28, 31); the filled points lost 0.1\%, 34\% and 44\% of their requests, unevenly by context but without relation to the column effect. The shaded band is the detection threshold ( dashed 0.01, dotted 0.05).}]{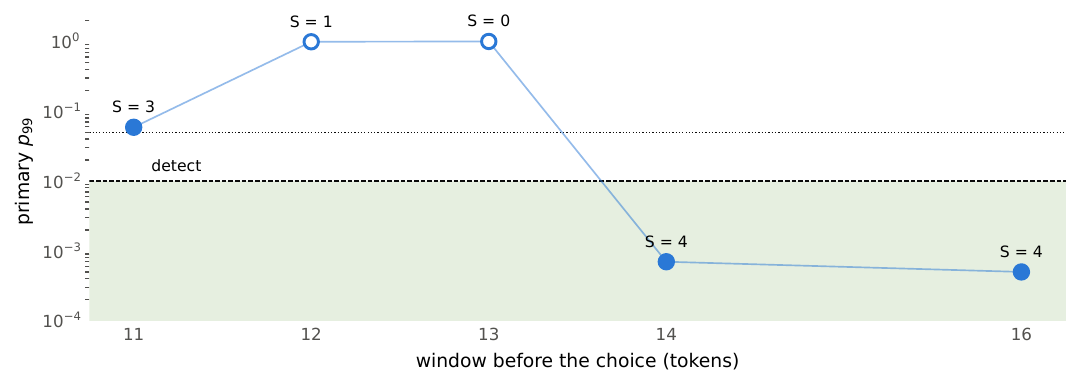}
\caption{The context-length search: primary \(p_{99}\) against the
window before the choice. Hollow points are the grids that the safety
classifier depleted unevenly (\(D = 28\), \(31\)); the filled points
lost 0.1\%, 34\% and 44\% of their requests, unevenly by context but
without relation to the column effect. The shaded band is the detection
threshold ( dashed 0.01, dotted 0.05).}\label{fig:8}
\end{figure}

The primary rejects at \(D=34\) and \(40\). This gives only a
conditional upper bound of fourteen tokens: the shift must be a hash of
preceding tokens, segmentation must follow Appendix~\ref{app:B}, and the
choice must follow the trailing space. There is no established lower
bound. The \(D=28\) and \(31\) null grids fail the refusal condition;
alternative base numbers were 84\% and 96\% refused and yielded no grid.
At \(D=25\) the primary is just outside the planned grey zone
(\(p_{99}=0.058\) initially, 0.059 pooled). Refusals at \(D=34\) are
also uneven, but uncorrelated with the shift (\(r=0.07\)); its
low-refusal cells retain range 0.56, \(p_{99}=0.0005\).

Secondary scores suggest another reading but cannot decide under the
plan. The paper's estimator rejects at \(D=25\) and \(28\); the range
rejects at \(D=25\), \(28\), \(34\) (0.40, 0.37, 0.51;
\(p_{99}\leq0.0011\)), but not \(31\) (0.20, \(p_{99}=0.16\)). A
fourteen-token window with shorter-length noise and a shorter window
obscured by refusals both fit. The latter would explain the aligned
effect at \(D=25\), where a hash longer than eleven tokens reaches the
prefix and cannot align across rows. Depleted grids also inflate the
code's \(\hat\sigma\) (1.6 on the second-base grid versus 1.1 in
Table~\ref{tab:15}), weakening the primary. Each length changes the hash
inputs, so detection need not be monotone once the window fits. The
stopping rule was not met; we do not select either reading.

\subsection{Mechanistic interpretations}\label{app:E.1}

The echo-versus-assembled contrast is consistent with prompt-aware
masking but does not identify the mechanism. Masking alone cannot
explain the nulls: P1, P24 and P40 omit the compact context from the
prompt and still fail. The account also requires sufficient context
length and a score tolerant of prefix effects. P40 fills a sixteen-token
window but remains null at ten to thirteen samples per cell with half
the prefixes nearly fixed on one word; the resampling comparison in
Table~\ref{tab:19} puts primary detection at about a half at that size.
It is consistent with the account without establishing it.

If the shift is a watermark, it excludes an inactive route and a fresh
independent key per request; a small key pool can retain average bias.
The length search then constrains the hash window conditionally. Without
that attribution, these remain explanations rather than identified
deployment properties. A provider detector could establish whether
responses carry a mark, but not whether marking caused the measured
shift.

\subsection{\texorpdfstring{The November
re-check}{The November re-check}}\label{app:E.2}

The plan \texttt{prereg\_W40b\_recheck.md}, written
on 8 November 2026 before collection, fixes the design of
Table~\ref{tab:7} to that of the pre-registered run: the same base
number, spaced form, list, nonce, batch transport and over-request
factor; ten prefixes at fifty planned samples per cell on Haiku 4.5 and
eight prefixes at twenty-five on Sonnet 5, with the within-prefix
code-estimator score at 10 000 permutations as primary, and the paper's
estimator, the published null, the continuous range and the verbatim
restriction as secondary. Its decision rules per model are: ``shift
appears'' if the primary \(p_{99} < 0.01\), the grid is nondegenerate,
refusals are not concentrated by context and the range is at least 0.3;
``shift absent'' if the primary \(p_{99} > 0.05\) and the range is below
0.2; anything else inconclusive. The correlation with Fable's September
columns is descriptive. Haiku 5.5 and Sonnet 5.5 were named exploratory.
Pilots of thirty-six requests per model preceded the plan and drew no
refusals; the three models of the Claude 5 generation accept no sampling
parameters, so provider defaults were used, as in September for Sonnet
5.

No request on any grid was refused. Parsing
succeeded for 2 299 of 2 304 Sonnet 5, 5 843 of 5 850 Haiku 4.5, 2 302
of 2 304 Sonnet 5.5 and 4 774 of 6 955 Haiku 5.5 responses; Haiku 5.5
rewrites its answer after ``Wait'' in about a third of responses, which
the parser rejects, and three further batches were submitted to fill its
cells. Verbatim copies of the number were 83\%, 46\%, 99\% and 75\%
(Haiku 4.5: 45\% in September); mean per-cell top shares 0.64, 0.44,
0.53 and 0.76. The pooled most frequent word is kiwis on Sonnet 5, Haiku
4.5 (September), Haiku 5.5 and Fable, dates on Haiku 4.5 (November) and
limes on Sonnet 5.5, whose kiwis columns are reported for the
correlation. On Sonnet 5 the published grid-wide score gives
\(p_{99} = 0.003\), the limes column means correlate with Fable's at
\(+0.97\), and the column means by final digit are \(+0.20\), \(+0.08\),
\(-0.24\), \(-0.21\), \(-0.27\), \(-0.05\), \(+0.21\), \(+0.16\),
\(+0.12\) for digits 1 to 9, against Fable's \(+0.23\), \(+0.27\),
\(-0.27\), \(-0.07\), \(-0.30\), \(-0.22\), \(+0.16\), \(+0.09\),
\(+0.10\). Under the plan, Sonnet 5 is inconclusive (primary in the grey
zone; range 0.47), Haiku 4.5 reads as shift absent (primary 1.00, range
0.13), the exploratory Sonnet 5.5 meets the detection rule and the
exploratory Haiku 5.5 is inconclusive (primary 1.00 at top share 0.76,
range 0.39). Batch identifiers are appended to the plan;
\texttt{analysis/recheck.py} recomputes every number of this section
from the stored responses.

\section{Prompts and word lists}\label{app:F}

\emph{Echo probe} (Gloaguen et al., with the format example):

\begin{quote}
Complete the sentence ``\{\(t_1\)\} \{\(t_2\)\}'' using only and exactly
a random word from the list: {[}\(\Sigma\){]}. Answer in this specific
format: \{\(t_1\)\} \{\(t_2\)\} strawberries. (the example uses a
different fruit on purpose, you have to choose among {[}\(\Sigma\){]})
\end{quote}

\emph{Described probe}:

\begin{quote}
Write ``\{\(t_1\)\}'', a space, the number made of \{\(H\) in words\}
copies of the digit \{\(d\)\}, a space, and one word chosen at random
from the list: {[}\(\Sigma\){]}. Reply with exactly that one line,
ending in a period, and nothing else. Format: \{\(t_1\)\}
\textless number\textgreater{} \textless word\textgreater.
\end{quote}

\emph{Spaced probe} (Section~\ref{sec:4}; the digits of \(t_2\)
separated by spaces):

\begin{quote}
Complete the sentence ``\{\(t_1\)\} \{4 7 3 8 \ldots{} 6 \(d\)\}'' using
only and exactly a random word from the list: {[}\(\Sigma\){]}. The
number is shown with a space after every digit; in your answer write its
digits together with no spaces. Answer in this specific format:
\{\(t_1\)\} \textless number\textgreater{} strawberries. (the example
uses a different fruit on purpose, you have to choose among
{[}\(\Sigma\){]})
\end{quote}

\emph{Masking differential}, type A (\(d_1 = c\)) and type B
(\(d_1 \neq c\)):

\begin{quote}
Write two lines. Line one: ``\{\(t_1\)\}'', a space, the number made of
\{\(H\)\} copies of the digit \{\(d_1\)\}, a space, and a random fruit
from {[}\(\Sigma\){]}. Line two: ``\{\(t_1\)\}'', a space, the number
made of \{\(H\)\} copies of the digit \{\(c\)\}, a space, and a random
fruit from {[}\(\Sigma\){]}, chosen independently. Each line ends with a
period. Nothing else.
\end{quote}

The \texttt{mixed} variant replaces ``\{\(H\)\} copies of the digit
\{\(d_1\)\}'' on line one with ``three copies of the digit \{\(e\)\}
followed by \{\(H-3\)\} copies of the digit \{\(d_1\)\}''.

\emph{Preamble} (runs B and B2): ``First write one paragraph of about
\{80 \textbar{} 150\} words explaining how the gears on a bicycle let a
rider climb a hill. Do not mention any fruit or write any digits in that
paragraph. Then, on a new line, \ldots{}'' followed by the echo probe.

\emph{System prompt}: the line ``{[}request-id: \{nonce\}{]}'', then on
a new line ``You complete sentences exactly as instructed. Reply with
the completed sentence and nothing else: no preamble, no explanation, no
quotation marks.''

\emph{Prefixes} \(t_1\): I ate, I chose, I picked, I selected, I took, I
went for, I settled on, I got, I gathered, I harvested.

\emph{Word lists}:
peaches/\allowbreak{}plums/\allowbreak{}cherries/\allowbreak{}apricots;
apples/\allowbreak{}bananas/\allowbreak{}oranges/\allowbreak{}pears;
mangoes/\allowbreak{}pineapples/\allowbreak{}papayas/\allowbreak{}kiwis;
strawberries/\allowbreak{}blueberries/\allowbreak{}raspberries/\allowbreak{}blackberries;
grapes/\allowbreak{}melons/\allowbreak{}lemons/\allowbreak{}figs;
plums/\allowbreak{}limes/\allowbreak{}dates/\allowbreak{}kiwis.

\emph{Non-word contexts} (run W), by digit: 1 zorp blen kwid tral; 2
mivo dast rulk pheng; 3 quimb narf tesk voli; 4 flimp gora tazz nuvel; 5
darbo yelk strun pavi; 6 ubbet kazlo trisp mordu; 7 sorvin plek dumo
waskel; 8 nefra tulb kreno vipp; 9 yavil bosk prund telm.

\section{Refusals}\label{app:G}

Table~\ref{tab:16} lists the requests that ended with
\texttt{stop\_reason:\ refusal}, category \texttt{cyber}, by prompt
family on \texttt{claude-fable-5-1}.

\begin{table}[tbp]
\centering
\caption{Refusals on \protect\texttt{claude-fable-5-1} by
prompt family.}\label{tab:16}
\vspace{\baselineskip}
\begin{tabular}{@{}
  >{\raggedright\arraybackslash}p{(\linewidth - 4\tabcolsep) * \real{0.3333}}
  >{\raggedright\arraybackslash}p{(\linewidth - 4\tabcolsep) * \real{0.3333}}
  >{\raggedright\arraybackslash}p{(\linewidth - 4\tabcolsep) * \real{0.3333}}@{}}
\toprule
\begin{minipage}[b]{\linewidth}\raggedright
prompt family
\end{minipage} & \begin{minipage}[b]{\linewidth}\raggedright
wording
\end{minipage} & \begin{minipage}[b]{\linewidth}\raggedright
refused
\end{minipage} \\
\midrule
echo, spelled digits & ``Complete the sentence \ldots{}'' & 11 of 3 510
(0.3\%) \\
echo, synchronous & same & 5 of 450 (1.1\%) \\
described & ``the digit 7 written twelve times in a row'' & 76 of 720
(10.6\%) \\
two-line, same digit & ``the digit 7 written twelve times in a row'' & 8
of 8 probes \\
two-line, same digit & ``the number made of twelve copies of the digit
7'' & 0 of 936 \\
two-line, mixed & ``three copies of the digit 5 followed by fifteen
copies of the digit 3'' & 113 of 936 (12.1\%) \\
non-word contexts & ``Complete the sentence \ldots{}'' with
\texttt{zorp\ blen\ kwid\ tral} & 82 of 720 (11.4\%) \\
described, long & ``the number made of forty copies of the digit 7'' &
306 of 1 300 (23.5\%) \\
spaced, 40 digits & ``Complete the sentence \ldots{}'' with the digits
separated by spaces & 470 of 1 500 (31.3\%) \\
spaced, 40 digits, plums list & same, pre-registered run; concentrated
on two prefixes, uneven over contexts & 3 875 of 8 840 (43.8\%) \\
echo, 40 digits, plums list & same number written compactly in the
prompt (run E40) & 0 of 2 304 \\
spaced, 40 digits, second base number & pilot-selected base (19\% in the
32-request pilot) & 847 of 2 304 (36.8\%), 3 cancelled \\
spaced, 40 digits, triple varied inside & same, last digit fixed; the
triple 444 refused most & 479 of 2 400 (20\%) \\
spaced, 40 digits, triple varied outside & same & 208 of 2 400
(8.7\%) \\
spaced, 25 digits & same, bisection point & 6 of 4 608 (0.1\%) \\
spaced, 28 digits & same; one (prefix, number) cell refused every time &
3 400 of 7 104 (47.9\%) \\
spaced, 31 digits & same; three of nine numbers refused in most prefixes
& 4 092 of 7 776 (52.6\%) \\
spaced, 34 digits & same & 1 064 of 3 104 (34.3\%) \\
spaced, 28 and 31 digits, second base number & same & 4 334 of 4 500
(96\%); 3 777 of 4 500 (84\%) \\
\bottomrule
\end{tabular}
\end{table}

A refused request returns \texttt{stop\_reason:\ refusal} with empty
text. That the decision is made on the prompt and not on the output is
indicated, not proved, by the echoed grid E40: the same forty digits and
the same answers drew no refusals written compactly and 44\% written
with spaces (Section~\ref{sec:4}). First-round refusals by cell are
reconstructed from the \texttt{attempt} field of the logs by
\texttt{analysis/refusals.py}. In the runs of Tables
13 and 15 the refusals fell on every context, unevenly
(Section~\ref{sec:4}); at \(D = 28\) and \(31\) of Table~\ref{tab:6}
whole contexts were refused. No refusals were observed on
\texttt{claude-haiku-4-5} or on \texttt{claude-sonnet-5}, nor on any of
the four November re-check grids (Section~\ref{sec:4.4}).

\section{Pilot grids and design checks}\label{app:H}

Before the pre-registered run of Section~\ref{sec:4} we ran the
long-context designs at ten to thirteen samples per cell. Runs P24 and
P40 extend the described probe to twenty-four and forty copies of one
digit (windows of ten and sixteen tokens); run W40 is the spaced design
of Section~\ref{sec:4} with the apples list; Haiku was run on both
designs as the control. Besides the published and within-prefix scores,
each grid was scored by the sum over prefixes of the chi-square of the
digit-by-word table, with the null relabelling the digit of every
response inside its prefix. That score asks only whether the choice
depends on the digit within a prefix and needs no agreement across
prefixes.

\begin{table}[tbp]
\centering
\caption{Longer contexts on Claude. \(p\) is the 99\% upper
bound; the published and the within-prefix columns give the code's and
then the paper's estimator, the former being the primary score of
Section~\ref{sec:4}, under which every pilot grid is at \(S \leq 2\); 10
000 permutations, scored as in Section~\ref{sec:4}.}\label{tab:17}
\vspace{\baselineskip}
\small\setlength{\tabcolsep}{3pt}
\begin{tabular}{@{}
  >{\raggedright\arraybackslash}p{(\linewidth - 16\tabcolsep) * \real{0.0538}}
  >{\raggedright\arraybackslash}p{(\linewidth - 16\tabcolsep) * \real{0.0692}}
  >{\raggedright\arraybackslash}p{(\linewidth - 16\tabcolsep) * \real{0.1846}}
  >{\raggedright\arraybackslash}p{(\linewidth - 16\tabcolsep) * \real{0.0923}}
  >{\raggedright\arraybackslash}p{(\linewidth - 16\tabcolsep) * \real{0.0538}}
  >{\raggedright\arraybackslash}p{(\linewidth - 16\tabcolsep) * \real{0.0769}}
  >{\raggedright\arraybackslash}p{(\linewidth - 16\tabcolsep) * \real{0.1231}}
  >{\raggedright\arraybackslash}p{(\linewidth - 16\tabcolsep) * \real{0.1923}}
  >{\raggedright\arraybackslash}p{(\linewidth - 16\tabcolsep) * \real{0.1538}}@{}}
\toprule
\begin{minipage}[b]{\linewidth}\raggedright
run
\end{minipage} & \begin{minipage}[b]{\linewidth}\raggedright
model
\end{minipage} & \begin{minipage}[b]{\linewidth}\raggedright
context
\end{minipage} & \begin{minipage}[b]{\linewidth}\raggedright
list
\end{minipage} & \begin{minipage}[b]{\linewidth}\raggedright
\(n\)
\end{minipage} & \begin{minipage}[b]{\linewidth}\raggedright
refused
\end{minipage} & \begin{minipage}[b]{\linewidth}\raggedright
published \(p\)
\end{minipage} & \begin{minipage}[b]{\linewidth}\raggedright
within-prefix \(S\), \(p\)
\end{minipage} & \begin{minipage}[b]{\linewidth}\raggedright
row \(\chi^2\), \(p\)
\end{minipage} \\
\midrule
P24 & Fable & described, 24 copies of one digit & apples & ≤ 13 & 22\% &
1.00 / 0.64 & 0, 1.00 / 2, 0.79 & 0.001 \\
P40 & Fable & described, 40 copies of one digit & apples & ≤ 13 & 23\% &
1.00 / 0.63 & 0, 1.00 / 3, 0.086 & 0.005 \\
W40 & Fable & spaced, 40 digits, last digit varies & apples & ≤ 13 &
31\% & 1.00 / 0.045 & 0, 1.00 / 3, 0.23 & \textbf{\boldmath 0.001} \\
--- & Haiku & described, 40 copies of one digit & apples & 13 & 0 & 1.00
/ 0.93 & 0, 1.00 / 1, 1.00 & 0.001 \\
--- & Haiku & spaced, 40 digits, last digit varies & apples & 13 & 0 &
0.73 / 0.79 & 2, 0.94 / 2, 0.96 & 0.75 \\
--- & Haiku & spaced, 40 digits, last digit varies & peaches & 13 & 0 &
0.73 / 0.77 & 1, 1.00 / 1, 1.00 & 0.21 \\
\bottomrule
\end{tabular}
\end{table}

The column statistics are null in every row but one: W40 rejects under
the published score with the paper's estimator (\(p_{99} = 0.045\)) and
under nothing else. The row chi-square rejects on Fable in all three
runs, but also on Haiku with the described probe, where the digit is
named in the instruction: an unwatermarked Claude ties its choice to a
salient digit, as the unwatermarked Qwen did in Section~\ref{sec:3}. The
same holds at full sample size: on the Haiku control of
Table~\ref{tab:15} the row chi-square rejects at \(p_{99} = 0.0005\)
while every column score is null, so an unwatermarked Claude does tie
its choice to the last digit within a prefix, in directions that do not
line up across prefixes. A within-prefix dependence on the digit is
therefore not evidence of a watermark on this family, and it is not used
in the paper. The continuous range statistic of Section~\ref{sec:4} is
exposed to the same reaction when it is aligned across prefixes: on the
Haiku described grid it gives 0.42, \(p_{99} = 0.0007\), and on the
Fable described grids 0.27 and 0.26 (\(p_{99} \leq 0.006\)); on the
spaced grids, where nothing in the context is conspicuous, it gives 0.12
and 0.13 on Haiku (\(p_{99} = 0.33\), \(0.11\)) and 0.36
(\(p_{99} = 0.010\)) on Fable with the apples list. All on the eight
prefixes of Table~\ref{tab:6}. The statistic is safe only on a design
where the contexts differ in nothing the model can notice, which is why
it is read on the last-digit grids and not on the described or
triple-varied ones. (On the echoed grid E40 the row chi-square reaches
only \(p_{99} = 0.07\).)

\textbf{\boldmath Follow-up design details.}

Table~\ref{tab:5} reports the four follow-up grids, each covered by a
plan addendum, using eight well-sampled prefixes and twenty-five planned
samples per cell. Here we give design details and checks beyond that
table.

\emph{E40.} The compact number replaces the spaced one in the prompt;
the number, list, window and score are otherwise unchanged. Each cell
received thirty-two responses. Both published scores give
\(p_{99}=1.00\); column correlation with W40b is \(+0.53\) over nine
columns, not significant.

\emph{W40c.} Changing the first thirty-nine digits changes twelve tokens
of a fourteen-token window while preserving the last digit. A pilot
chose the least-refused of eight bases. Conditional on primary
detection, the plan interprets column correlation with W40b below 0.3 as
a replicated keyed shift and above 0.7 as a native preference. Refusals
depleted four prefixes and emptied one cell. On the four complete
prefixes, the paper's estimator drops to \(S=2\); column correlation is
\(-0.19\), versus \(-0.34\) over all eight. Digits 3 and 5 now favour
kiwis, and 7 is the strongest column. Because the primary does not
reject, its correlation criterion cannot decide. The range statistic was
not named in this addendum and is exploratory. Two further pilots over
twenty-four bases found none refused below 19\%.

\emph{Inside and outside.} Both fix the final digit at 4, W40b's
highest-entropy column, and vary \(ddd\), \(d=1,\dots,9\), at positions
28--30 or 1--3: the sixth or fifteenth token before the choice. Token
counts support unchanged boundaries (Appendix~\ref{app:B}). The plan
makes the within-prefix score primary and the range secondary. Both
grids fail the refusal condition: inside, triple 444 is
disproportionately refused; outside, refusals are also uneven despite a
lower overall rate. The inside range has bootstrap interval 0.49 to 0.63
and spans columns; outside it is 0.18 to 0.31 and is carried by 111,
which moves the choice from limes to kiwis and dates. Such
conspicuous-string effects also occur on the unwatermarked described
grids above, so these secondary scores cannot identify a hash.

\textbf{\boldmath Design checks.}

Table~\ref{tab:18} compares the published and within-prefix scores on
reference grids from Table~\ref{tab:1} and new last-digit-varying
contexts. Both detect the stock scheme and remain blind to the
prompt-aware variant, which masks echoed contexts. The code's estimator
stays clean on the control. Unlike repeated-digit runs, the new contexts
do not repeat internally, so stock detection remains maximal at \(H=8\)
as well as \(H=4\).

The described probe detects the prompt-aware reference at \(S=8\)
(Table~\ref{tab:1}), demonstrating the benefit of constructing the
context. The final spaced form achieves this construction in 99.7\% of
Fable responses, but the 3B reference drops the number often enough to
leave four to thirteen of ninety cells empty. Its scorer locates the
choice by the last digit run without verifying the requested number
(Appendix~\ref{app:A.4}). On the remaining cells, the spaced probe at
\(H=4\) and \(H=8\) detects stock SynthID under the code's estimator
(\(S=4\), \(p=0.016\); \(S=5\), \(p=0.0005\)), stays clean on the
control (\(p=0.69\), \(1.00\)), and remains blind to the prompt-aware
variant (\(S=0\)). On the 3B reference the final
combination of the spaced probe and within-prefix score therefore lacked
a positive control on a prompt-aware reference. Read with the continuous
statistic, the same grids already contained it: the prompt-aware
variant's range is 0.88 and 0.77 at \(H = 4\) and 8 against 0.84 and
0.82 for the stock variant and 0.20 and 0.33 for the control, hidden
from \(S\) by the empty cells. Section~\ref{sec:3.3} repeats the checks
on Qwen3-8B-Instruct (4-bit, \texttt{unsloth} weights) with every
response verified to contain the requested number and regenerated
otherwise (twelve, eight and four of ninety at forty digits for the
three variants; one at eight): the spaced probe detects the stock and
the prompt-aware variant identically at \(H = 4\), 8 and 40, the echo
probe is masked under the prompt-aware variant at \(H = 40\), and the
unwatermarked control's range grows with the number length (0.36, 0.41,
0.59). Resampling that control at fifty draws per cell, the
within-prefix primary rejects in five of five draws (\(S = 3\) to 6),
while the two watermark variants give \(S = 7\) to 9. Data and scripts:
\texttt{runs/qwen8b/}, \texttt{analysis/local\_probe\_stats.py}.

\begin{table}[tbp]
\centering
\caption{The last-digit-varying contexts and the within-prefix
score on the reference model (Qwen2.5-3B, \(h = 4\), ten samples per
cell, code's \(\hat\sigma\); the one-digit rows are the grids of
Table~\ref{tab:1}; 10 000 permutations). At \(H = 8\) the paper's
estimator flags the control at \(S = 4\), \(p = 0.009\); the code's does
not.}\label{tab:18}
\vspace{\baselineskip}
\begin{tabular}{@{}
  >{\raggedright\arraybackslash}p{(\linewidth - 8\tabcolsep) * \real{0.2621}}
  >{\raggedright\arraybackslash}p{(\linewidth - 8\tabcolsep) * \real{0.0680}}
  >{\raggedright\arraybackslash}p{(\linewidth - 8\tabcolsep) * \real{0.2427}}
  >{\raggedright\arraybackslash}p{(\linewidth - 8\tabcolsep) * \real{0.2136}}
  >{\raggedright\arraybackslash}p{(\linewidth - 8\tabcolsep) * \real{0.2136}}@{}}
\toprule
\begin{minipage}[b]{\linewidth}\raggedright
probe, contexts, score
\end{minipage} & \begin{minipage}[b]{\linewidth}\raggedright
\(H\)
\end{minipage} & \begin{minipage}[b]{\linewidth}\raggedright
stock SynthID
\end{minipage} & \begin{minipage}[b]{\linewidth}\raggedright
control
\end{minipage} & \begin{minipage}[b]{\linewidth}\raggedright
prompt-aware
\end{minipage} \\
\midrule
echo, one digit repeated, published & 4 & \(S = 9\), \(p = 0.0005\) &
\(S = 1\), \(p = 1.00\) & \(S = 1\), \(p = 1.00\) \\
echo, one digit repeated, within-prefix & 4 & \(S = 9\), \(p = 0.0005\)
& \(S = 1\), \(p = 1.00\) & \(S = 1\), \(p = 1.00\) \\
echo, last digit varies, published & 4 & \(S = 10\), \(p = 0.0005\) &
\(S = 1\), \(p = 1.00\) & \(S = 1\), \(p = 1.00\) \\
echo, last digit varies, within-prefix & 4 & \(S = 10\), \(p = 0.0005\)
& \(S = 1\), \(p = 1.00\) & \(S = 1\), \(p = 1.00\) \\
echo, last digit varies, published & 8 & \(S = 10\), \(p = 0.0005\) &
\(S = 1\), \(p = 1.00\) & \(S = 1\), \(p = 1.00\) \\
echo, last digit varies, within-prefix & 8 & \(S = 10\), \(p = 0.0005\)
& \(S = 1\), \(p = 1.00\) & \(S = 1\), \(p = 1.00\) \\
\bottomrule
\end{tabular}
\end{table}

Second, the within-prefix null. Table~\ref{tab:19} resamples the
pre-registered Fable and Haiku grids of Table~\ref{tab:15} at smaller
sample sizes and scores every draw under both nulls. On Fable, whose
choice depends strongly on the prefix, the within-prefix null detects
more often than the grid-wide one at every sample size; on Haiku both
stay at or below the nominal false-positive rate. The draws come from a
grid that rejects, so the Fable columns are not an independent test;
they compare the power of the two nulls on the same data at reduced
size, and the Haiku columns are a calibration check on one grid, not a
false-positive rate.

\begin{table}[tbp]
\centering
\caption{Rate of \(p_{99} < 0.05\) over one hundred bootstrap
draws (with replacement) from the grids of Table~\ref{tab:15}, eight
prefixes, code's \(\hat\sigma\), 1 500 permutations per draw; the
threshold is 0.05, not the plan's 0.01.}\label{tab:19}
\vspace{\baselineskip}
\begin{tabular}{@{}
  >{\raggedright\arraybackslash}p{(\linewidth - 8\tabcolsep) * \real{0.1515}}
  >{\raggedright\arraybackslash}p{(\linewidth - 8\tabcolsep) * \real{0.1919}}
  >{\raggedright\arraybackslash}p{(\linewidth - 8\tabcolsep) * \real{0.2323}}
  >{\raggedright\arraybackslash}p{(\linewidth - 8\tabcolsep) * \real{0.1919}}
  >{\raggedright\arraybackslash}p{(\linewidth - 8\tabcolsep) * \real{0.2323}}@{}}
\toprule
\begin{minipage}[b]{\linewidth}\raggedright
\(n\) per cell
\end{minipage} & \begin{minipage}[b]{\linewidth}\raggedright
Fable, grid-wide
\end{minipage} & \begin{minipage}[b]{\linewidth}\raggedright
Fable, within-prefix
\end{minipage} & \begin{minipage}[b]{\linewidth}\raggedright
Haiku, grid-wide
\end{minipage} & \begin{minipage}[b]{\linewidth}\raggedright
Haiku, within-prefix
\end{minipage} \\
\midrule
5 & 0.35 & 0.55 & 0.02 & 0.01 \\
10 & 0.52 & 0.57 & 0.00 & 0.00 \\
25 & 0.43 & 0.69 & 0.01 & 0.00 \\
\bottomrule
\end{tabular}
\end{table}

\section{Reproduction and deviations}\label{app:I}

The repository contains the implementation, every parsed response
(refusals, request nonces and usage metadata are not logged), and the
scripts that produce every table and figure. The local experiments of
Section~\ref{sec:3} need one consumer GPU and no credentials. The hosted
runs of Section~\ref{sec:4} need an Anthropic API key. Each hosted run
is reproducible from its stored matrix without re-querying.

The reference statistic is the released code of Gloaguen et
al.~(\url{https://github.com/eth-sri/watermark-detection},
\texttt{src/attacks/redgreen\_test.py}), which the test suite fetches
and compares against our implementation; the reference watermark is
\texttt{SynthIDTextWatermarkLogitsProcessor} of \texttt{transformers}
5.16.1. The anonymous supplementary archive freezes the current scoring
scripts and stored runs, including \texttt{analysis/equal\_n.py}, with
per-file SHA-256 hashes. The scorer's published rows were edited after
collection (one pseudo-count per word, a common fill rule), which moves
some published \(p\)-values from the run logs but no primary one. The
pre-registration plans are kept in the \texttt{runs} directory of the
repository as two Markdown files. The Batches API assigns every
submission an identifier whose server-side creation time the provider
holds; the identifiers of every pre-registered batch are listed at the
end of the two plan files.

Deviations from the plans: at \(D = 25\) the first round gave
\(p_{99} = 0.058\), just outside the plan's grey zone
(\(0.01 \leq p_{99} \leq 0.05\), where a second round is collected), and
a second round was collected anyway; second rounds were also collected
at \(D = 28\) and \(31\), where the first rounds were clean nulls,
because both grids had failed the refusal condition; the refusal
condition of the W40b plan is not met (Section~\ref{sec:4}); the Haiku
control received 65 samples per cell and the later grids 32, because the
batch transport over-requests by 30\%; the continuous statistic was
named as secondary only from the sixth addendum on, and is exploratory
on the earlier grids; the bisection's stopping rule (both \(D_{\min}\)
and \(D_{\min} - 3\) decided on grids that meet the refusal condition)
was not met; the bisection ran without a control per point; the
second-base replication and the inside grid of the window check were
scored although their grids failed the refusal condition; the echo
counterpart E40 and the replication used the eight well-sampled prefixes
rather than ten. The re-check plan
(Appendix~\ref{app:E.2}) was followed; its one deviation is the three
extra Haiku 5.5 batches submitted to fill cells depleted by unparsable
responses.

\end{document}